%% file: 0_main.tex
\documentclass[lettersize,journal]{IEEEtran}

\ifCLASSOPTIONcompsoc
  \usepackage[nocompress]{cite}
\else
  \usepackage{cite}
\fi
\ifCLASSINFOpdf
\else
\fi
\usepackage{amsmath}
\usepackage{array}

\usepackage{times}
\usepackage{epsfig}
\usepackage{graphicx}
\usepackage{amssymb}

\usepackage{appendix}
\usepackage{bm} 
\usepackage{multirow}
\usepackage{float}

\usepackage{epsfig}
\usepackage{xcolor}
\usepackage[linesnumbered,ruled]{algorithm2e}
\usepackage[binary-units]{siunitx}
\usepackage{booktabs}
\usepackage{mathtools}
\usepackage{vcell}
\usepackage{enumitem}
\usepackage{subcaption}
\usepackage{tikz}
\usetikzlibrary{calc}

\usepackage{hyperref}
\hypersetup{
    colorlinks=true,
    linkcolor=blue,
    filecolor=magenta,      
    urlcolor=cyan,
}

\newcommand\todo[1]{\textcolor{red}{#1}}

\usepackage{circledsteps}

\begin{document}
%
\title{Sign-Coded Exposure Sensing for Noise-Robust High-Speed Imaging}

\author{R.~Wes~Baldwin,
        ~Vijayan~Asari,~\IEEEmembership{Senior~Member,~IEEE}
        and ~Keigo~Hirakawa,~\IEEEmembership{Senior~Member,~IEEE}

\thanks{R. Wes Baldwin is with Riverside Research, Fairborn, OH 45324 USA (e-mail: rwbaldwin@riversideresearch.org)}

\thanks{V. Asari, and K. Hirakawa are with the Department of Electrical and Computer Engineering, University of Dayton, Dayton, OH 45469 USA (e-mail: vasari1@udayton.edu; khirakawa1@udayton.edu)}
}

%
%

\markboth{Journal of \LaTeX\ Class Files,~Vol.~14, No.~8, August~2015}%
{Shell \MakeLowercase{\textit{et al.}}: Bare Demo of IEEEtran.cls for Computer Society Journals}
%



\IEEEtitleabstractindextext{%
\begin{abstract}
We present a novel Fourier camera, an in-hardware optical compression of high-speed frames employing pixel-level \emph{sign-coded exposure} where pixel intensities temporally modulated as positive and negative exposure are combined to yield Hadamard coefficients. The orthogonality of Walsh functions ensures that the noise is not amplified during high-speed frame reconstruction, making it a much more attractive option for coded exposure systems aimed at very high frame rate operation. Frame reconstruction is carried out by a single-pass demosaicking of the spatially multiplexed Walsh functions in a lattice arrangement, significantly reducing the computational complexity. The simulation prototype confirms the improved robustness to noise compared to the binary-coded exposure patterns, such as one-hot encoding and pseudo-random encoding. Our hardware prototype demonstrated the reconstruction of 4kHz frames of a moving scene lit by ambient light only.
\end{abstract}

}

\maketitle

\IEEEdisplaynontitleabstractindextext

%
\IEEEpeerreviewmaketitle

\input{1_intro}

\input{2_priorWork}

\input{3_principle}

\input{4_design}

\input{5_challenges}

\input{6_experiments}

\input{7_conclusion}

\ifCLASSOPTIONcaptionsoff
  \newpage
\fi



%
{\small
\bibliographystyle{IEEEtran}
\bibliography{egbib}
}
%
%

%





\begin{IEEEbiography}[{\includegraphics[width=1in,height=1.25in,clip,keepaspectratio]{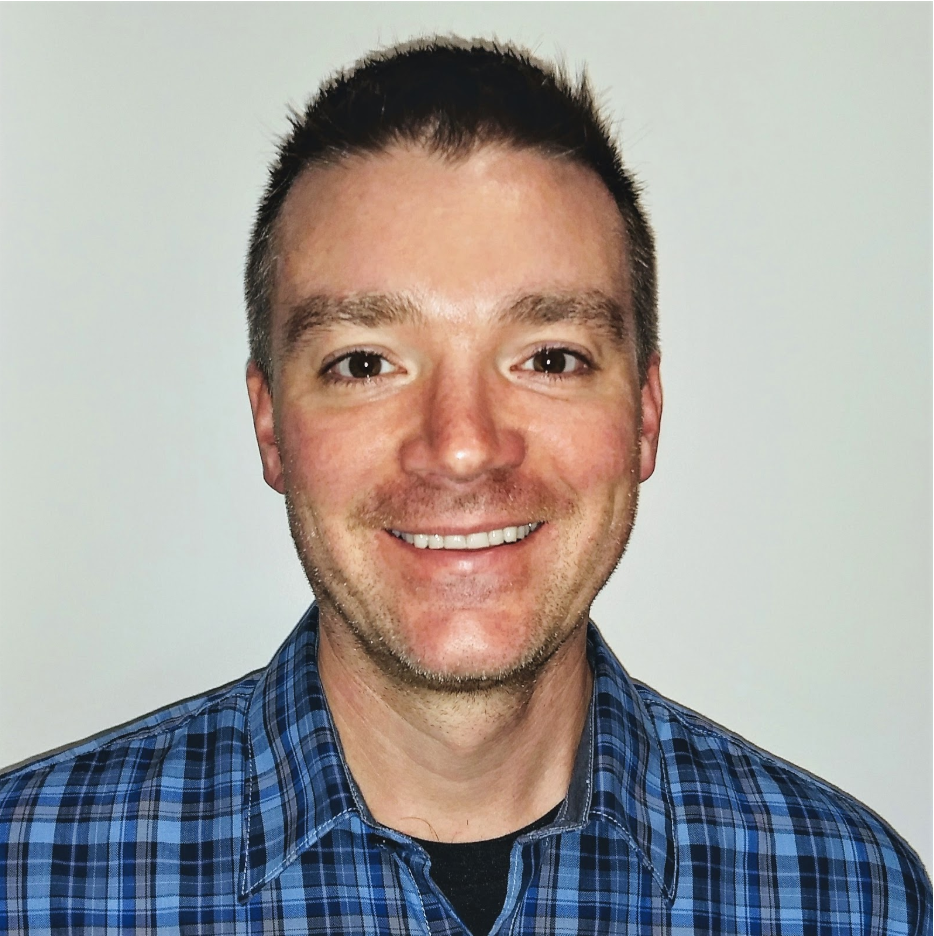}}]{R. Wes Baldwin}
received a B.S. in Computer Engineering in 2002 from Kettering University, a M.S. in Electrical and Computer Engineering in 2005 from the University of Illinois at Chicago, and a Ph.D. in Electrical and Computer Engineering from the University of Dayton in 2021. He has over 15 years of experience working as a research engineer for both the National Air and Space Intelligence Center and the National Geospatial-Intelligence Agency (NGA). He currently works as an AI/ML engineer for Riverside Research, an independent National Security Nonprofit. His research interests include remote sensing, machine learning, image processing, and event cameras.
\end{IEEEbiography}

\begin{IEEEbiography}[{\includegraphics[width=1in,height=1.25in,clip,keepaspectratio]{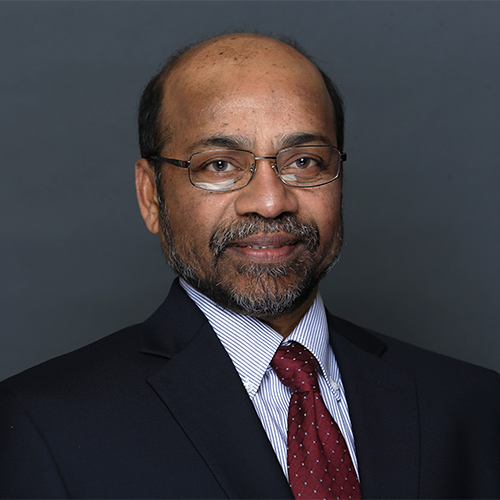}}]{Vijayan Asari}
Vijayan Asari received the Ph.D. degree in electrical engineering from the Indian Institute of Technology Madras in 1994. He is currently a Professor in electrical and computer engineering and Ohio Research Scholars Endowed Chair in wide area surveillance with the University of Dayton, Dayton, Ohio, USA. He is also the director of the Center of Excellence for Computational Intelligence and Machine Vision (Vision Lab) at UD. Dr. Asari holds four United States patents and has published more than 700 research articles including an edited book in wide area surveillance and 116 peer-reviewed journal papers in the areas of image processing, pattern recognition, machine learning, deep learning, and artificial neural networks. He is an elected Fellow of SPIE and a senior member of IEEE, and a co-organizer of several SPIE and IEEE conferences and workshops. Dr. Asari received several teaching, research, advising and technical leadership awards including the University of Dayton School of Engineering Vision Award for Excellence in August 2017, the Outstanding Engineers and Scientists Award for Technical Leadership from The Affiliate Societies Council of Dayton in April 2015, and the Sigma Xi George B. Noland Award for Outstanding Research in April 2016.
\end{IEEEbiography}

\begin{IEEEbiography}[{\includegraphics[width=1in,height=1.25in,clip,keepaspectratio]{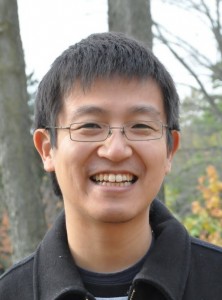}}]{Keigo Hirakawa}
(S’00–M’05–SM’11) received the
B.S. degree (Hons.) in electrical engineering from
Princeton University, Princeton, NJ, USA, in 2000,
the M.S. and Ph.D. degrees in electrical and computer engineering from Cornell University, Ithaca,
NY, USA, in 2003 and 2005, respectively, and the
M.M. degree (Hons.) in jazz performance studies
from the New England Conservatory of Music,
Boston, MA, USA, in 2006. He was a Research
Associate with Harvard University, Cambridge, MA,
USA from 2006 to 2009. He is currently an Associate Professor with the University of Dayton, Dayton, OH, USA. He is
currently the Head of the Intelligent Signal Systems Laboratory, University of
Dayton, where the group focuses on statistical signal processing, color image
processing, and computer vision.
\end{IEEEbiography}





\end{document}

%% file: 1_intro.tex
\section{Introduction}

\begin{figure*}
\begin{tabular}{ccc}
     \includegraphics[width=.31\linewidth]{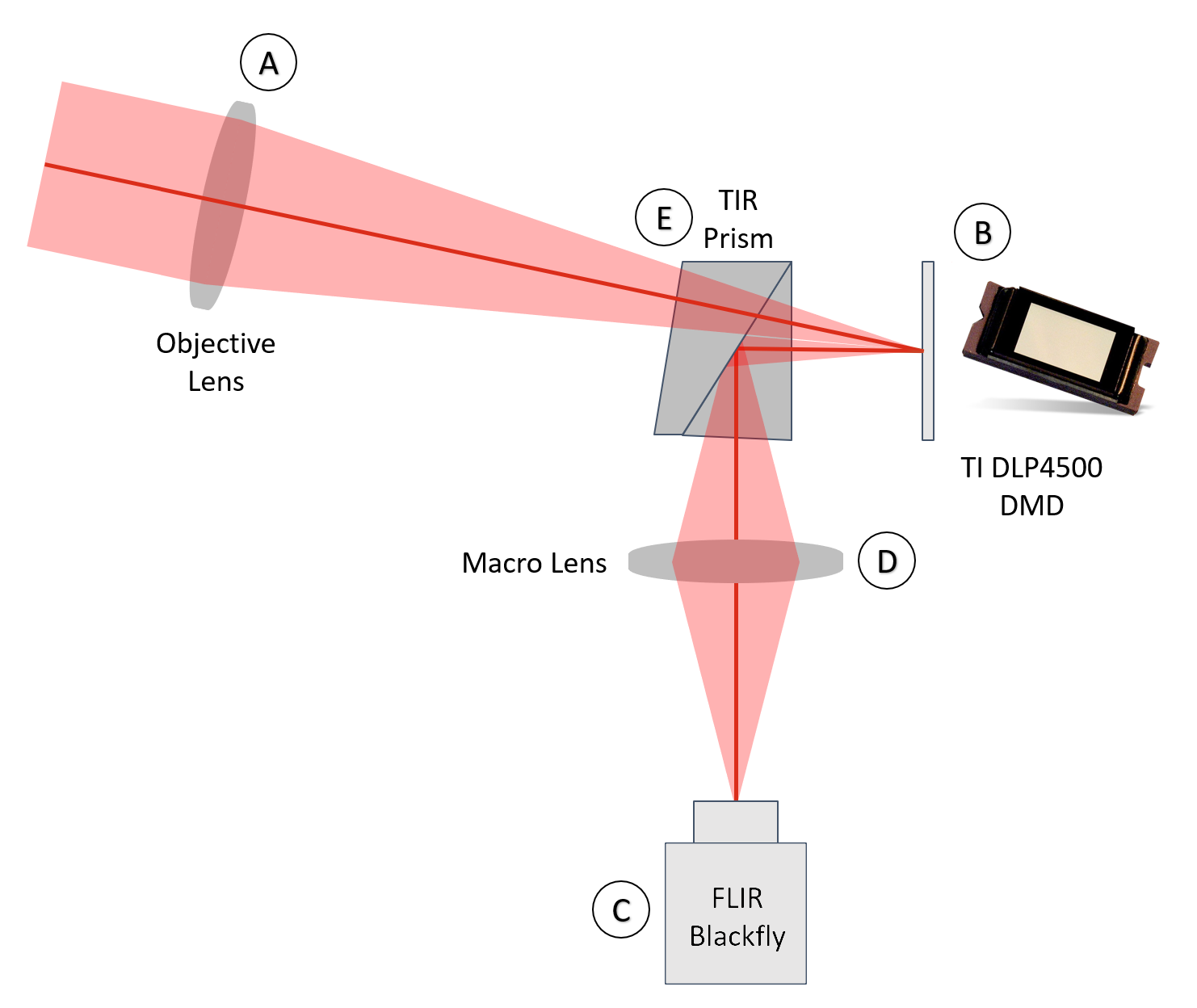}&
     \includegraphics[width=.31\linewidth]{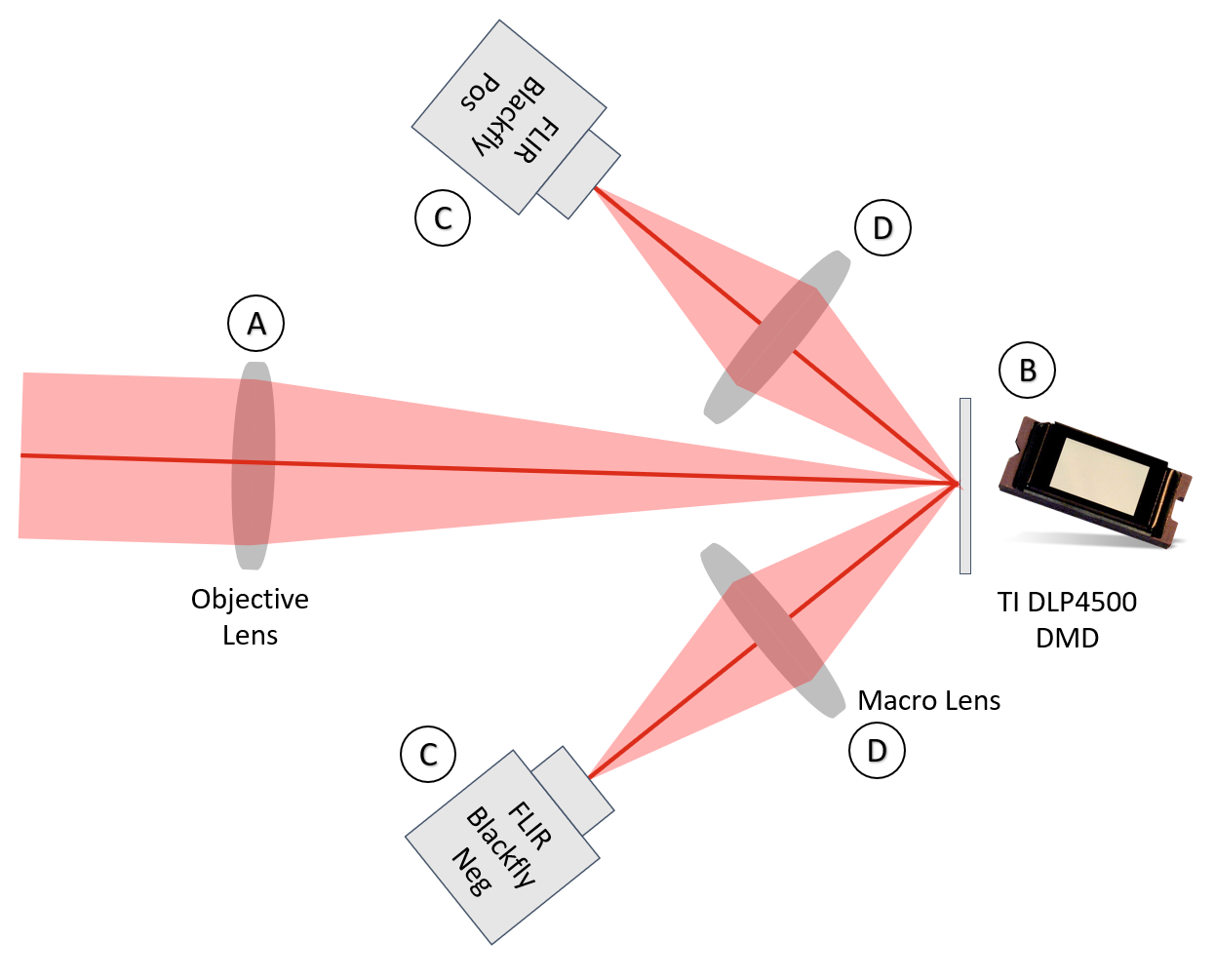}&
    \includegraphics[width=.31\linewidth]{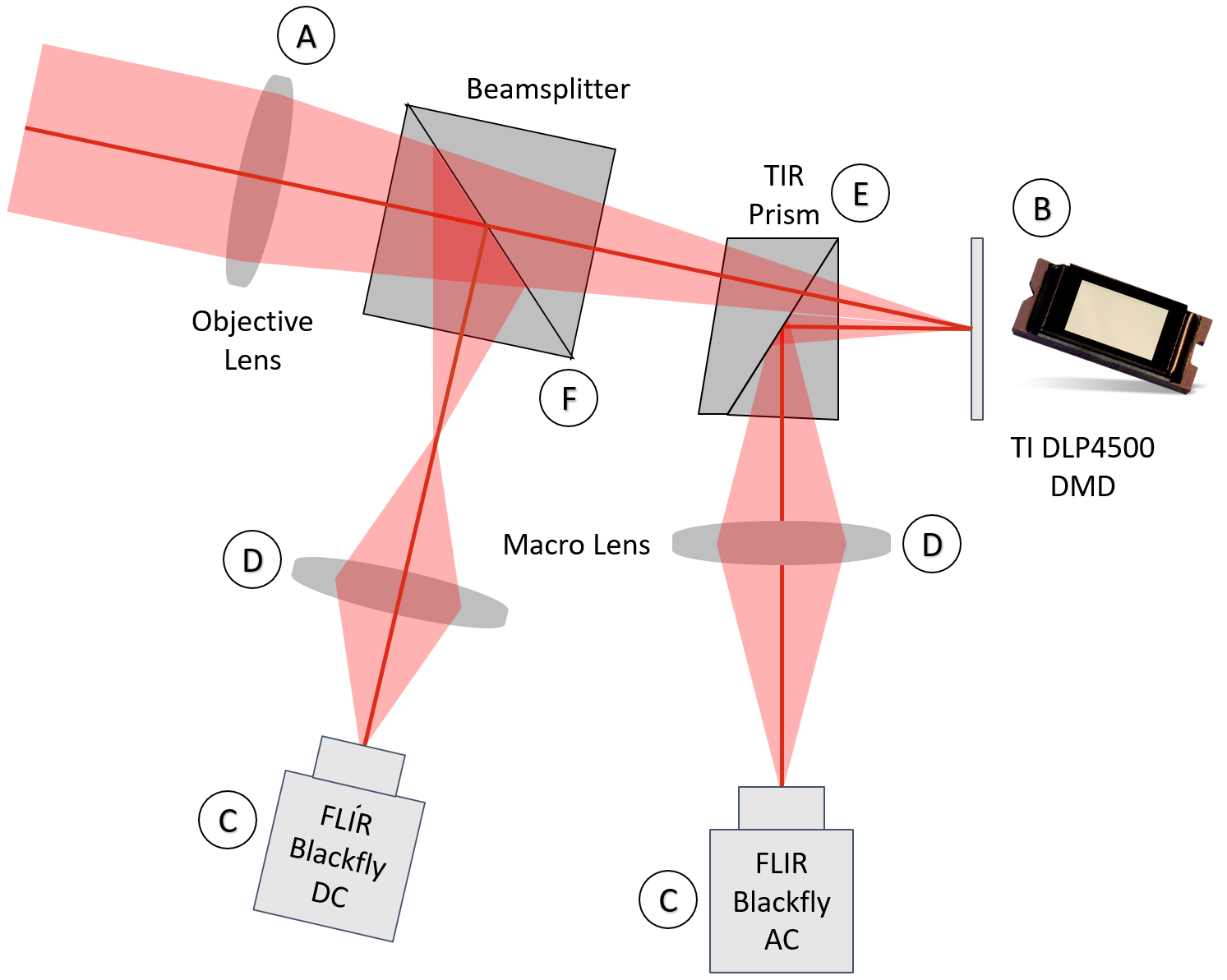}\\
    (a) Binary-Coded Exposure Camera&(b) Sign-Coded Fourier Camera&(c) Sign-Coded Fourier Camera\\&(Design \#1)&(Design \#2)\\
    \end{tabular}
    \caption{Binary-coded and sign-coded exposure cameras. The objective lens in \Circled{A} focuses the image onto DMD in \Circled{B}, which is observed by the camera(s) in \Circled{C} via relay/macro lens(es) in \Circled{D}. The total inner reflection prism in \Circled{E} directs 100\% of the DMD-modulated light towards the camera. The conventional {\bf binary-coded exposure camera} in (a) encodes the temporal evolution of the intensity by the DMD blocking or passing the light at each DMD pixel. Not only is this design inferior in terms of light efficiency, but the reconstruction is mathematically proven to amplify noise (for random binary modulation, condition number$>$100). By contrast, the proposed {\bf sign-coded exposure camera} design in (b) employs two cameras representing ``positive'' and ``negative'' light modulations. The light efficiency is 100\%, and reconstruction with no noise amplification is possible (i.e.~condition number=1). The alternative design in (c) employs an additional beam splitter in \Circled{F} to capture a non-modulated signal---sign-coded light modulation signal is computed in post-processing. This design is easier to implement, with only a small sacrifice in light efficiency and condition number.}
    \label{fig:camera_design}
\end{figure*}


Frame data throughput is a primary bottleneck in video processing. The imbalance between data quality and size is most evident in high-speed cameras where fast frame rates generate staggering amounts of data, even though there is relatively little new information introduced per frame. For this reason, high-speed cameras selectively reduce data volume via hardware-based cropping and temporal triggering. Modern video compression algorithms can significantly reduce data volume, but these algorithms rely on frame data as input---still requiring large data readout and processing from the sensor. 

In recent years, there have been efforts to develop \emph{in-hardware} compression to bypass large-data handling. It replaces redundant video frames with a single dense encoded image to eliminate the need to transfer or handle large data. Examples include compressive sensing-inspired coded exposure and event detection cameras. Unfortunately, in-hardware compression methods significantly alter the nature of the captured data, are sensitive to noise, and/or require computationally expensive algorithms to reconstruct frames. In particular, noise amplification is a major limiting factor for compressive sensing to operate at very high frame rates---an imaging modality where the sensors are already photon-starved.


We propose a novel coded exposure-based high-speed camera design that encodes multiple frames into a single image in real time that we call a Fourier Camera (FC). Rather than using random pixel encoding patterns, we propose using structured patterns in a lattice grid similar to a color filter array (CFA) in traditional cameras. Our method is computationally efficient and avoids solving large underdetermined systems through spatial-temporal demosaicking. 
The proposed design assists in capturing fast motion as well as fast intensity changes beyond the frame rate limitations of the camera.
We achieve 4kHz frame rate reconstruction with reasonable noise suppression---a rate we accomplish by explicitly considering noise robustness and reconstruction stability along with the properties of high-speed image signals. Additionally, frame reconstruction is extremely efficient and can easily be implemented to execute in real-time.

We summarize our contributions below:
\begin{itemize}[leftmargin=5.5mm]
  \item {\bf Sign-Coded Fourier Camera (FC):} We propose a new FC imaging architecture to encode positive and negative exposures in real-time in-hardware compression of video sequences at 4k frames per second (FPS)
  \item {\bf Noise Robustness:} We prove that Hadamard-based sign-coded exposures yield 100\% light efficiency and low or no noise amplification during reconstruction. Noise robustness is key to pushing the frame rate of high-speed imaging to photon-starved speeds.
  \item {\bf Temporal Demosaicking:} We test and implement several temporal demosaicking methods to maximize reconstruction accuracy. We simulate the design, evaluate performance, and assess noise robustness using multiple datasets.
  \item {\bf Prototype Hardware:} We demonstrate FC by building a hardware prototype and reconstructing sequences at 4k frames per second (FPS), limited only by hardware and not by the design of the coded exposure.
\end{itemize}

%% file: 2_priorWork.tex
\section{Prior Work}
\label{sec:priorwork}

\begin{table*}
    \centering
    \caption{Noise sensitivity as assessed by condition number of binary sensing matrices. The sensing matrix size is 16$\times$16. For binary random sensing matrix, exactly 50\% of the entries in each row are ``1.'' Because it is random, the condition number is also random. We generated 1 million random binary matrices and report the median condition number. (The average condition number is $\infty$ because not all random binary matrices are non-singular.) We also enumerate the reported speed at which the real hardware prototypes operate in the respective publications.}
    \label{tab:cond_number}
    \begin{tabular}{|c||c|c||c|c|c|c|}
\hline
    Sensing Matrix & Exposure Coding & \# Cameras & Condition Number& Light Efficiency &  Modulator Speed & Camera Speed \\
    \hline\hline
    One-Hot \cite{bub2010temporal}& Binary& 1&1 & 1/16 &400Hz&25Hz\\
    \hline
    Pseudo-Random\cite{reddy2011p2c2}& Binary &1&median=113.34 &1/2&184Hz&23Hz\\        
    \hline
    Positive Hadamard & Binary & 1& 9.90 & 1/2&not tested&not tested\\
    \hline
     Hadamard (Design \#1)& Signed& 2& 1 & 1&not tested&not tested\\
    \hline
    Hadamard (Design \#2) & Signed& 2& 2.6180 & 1/2+1/4&4000Hz&250Hz\\
\hline
    \end{tabular}
\end{table*}

\textbf{High-speed imaging} seeks to capture extremely high frame rate videos using custom hardware. Over the last several decades, this hardware has grown smaller, cheaper, and more complex. Many cellphones have now mainstreamed high-speed imaging by enabling video recording up to 240 FPS~\cite{lincoln2017enhancing}. Commercially available scientific hardware now easily captures high-definition video at over 25,000 FPS~\cite{jozwik2016industrial}, and the most specialized, custom-built laboratory cameras now image at over 70 trillion FPS~\cite{wang2020single}. The proposed FC is an alternative low-cost solution that achieves similar performance to commercially available high-speed cameras, designed to adjust temporal sampling patterns without a significant increase in the data volume.

\textbf{Video compression} uses motion compensation and inter-frame coding to eliminate redundant information and reduce overall data volume. This is particularly appealing to high-speed imaging, since the frames are highly correlated to each other. However, current compression approaches are software-based and require complete frame readout before compression. Large data readout and subsequent compression can act as a bottleneck due to high computational cost, high power consumption, and increased latency.

\textbf{Coded shutter} (a.k.a.~flutter shutter, coded exposure) is a mechanical device that modulates the intensity during the camera exposure, which allows object motion to be encoded by the recorded pixels and deblurred in post processing~\cite{raskar2006coded,holloway2012flutter}. Coded exposure has also been implemented using strobe light~\cite{kappal2009illustrating}. \textbf{Spatial light modulator} (SLM) devices such as a digital micromirror device (DMD) and liquid crystal on silicon (LCoS) have been used in the optical path of the imaging system to spatially and temporally modulate the pixel intensity simultaneously~\cite{nayar2004programmable}. Commonly used in projectors, SLM occurs at rates imperceptible to the human eye. Sometimes called ``DMD cameras,'' hardware configuration shown by Figure \ref{fig:camera_design}(a) has been used to generate high dynamic range (HDR) and high-speed images using standard sensors by masking out very bright sources or varying exposure per pixel or encoding the temporal evolution of the intensity at each pixel using a binary (on-off) pattern. More recently, sensor hardware to replace spatial light modulators have been developed~\cite{luo2019cmos}. This novel hardware allows pixel-level photoelectric integration to be temporarily modulated during a single detector readout cycle, achieving the same effect as DMD and LCoS without additional optical elements.

Leveraging \textbf{compressive sensing} principles, spatial-temporal light modulators such as coded shutter and DMD cameras have been used to compress video during data acquisition~\cite{holloway2012flutter,deng2019sinusoidal,liu2013efficient,mochizuki2016single,bub2010temporal,yuan2021snapshot,liu2018rank}. For example, linear motion encoded by coded shutter camera is modeled, which is used to recover the frames in post processing \cite{holloway2012flutter}. This idea can be extended to per-pixel coded exposure using spatial light modulators, where SLM encodes the spatial-temporal evolution of a moving scene by blocking or transmitting light at specific time instances forming a pseudo-random binary coded pattern during a single detector readout cycle. Subsequent processing reconstructs video frames from a single snapshot capture exposed over the per-pixel binary coded exposure movements using linear programming. This configuration allows video to be recorded at significantly lower bandwidth, provided that there exists an underlying representation of the video signal that is sparse\cite{reddy2011p2c2}. Alternatively, work in~\cite{bub2010temporal} proposed per pixel grid-based encoding strategy, activating the integration for a very short time per pixel (which we refer to as ``one-hot'' in our experiments). 

Compressive sensing is well-matched for high-speed video, as there is relatively little new information introduced per frame. Such assumptions promote sparse representation that can be exploited during frame reconstruction. However, noise sensitivity ultimately limits the achievable frame rate in high-speed imaging using compressive sensing approaches, since low photon count is inversely proportional to the frame rate. Besides the light efficiency, the empirical condition number of pseudo-random binary sensing matrix used in compressive sensing techniques is extremely high, as evidenced in Table~\ref{tab:cond_number}. This implies that a small perturbation in the measurement due to noise or calibration error will be amplified during the frame reconstruction, making it challenging to operate in photon-limited conditions. Alternatively, the one-hot encoding approach in~\cite{bub2010temporal} can reconstruct frames without noise amplification (i.e.~condition number is 1), but this approach severely limits the number of photons available to the detector, yielding low SNR images.

Although not directly related to our work, there are alternatives to reducing the data volume in high-speed video. \textbf{Video interpolation} is a technique to estimate intermediary frames based on still images or slow video sequences. For instance, video reconstruction methods in ~\cite{zakharov2019few,jin2018learning} use trained generative adversarial  networks~\cite{goodfellow2014generative} to generate plausible motion of a human head from a single input image. Another method uses deep learning to yield a high-speed video from standard video~\cite{jiang2018super} by generating intermediary frames that are temporally and spatially consistent with the two consecutive recorded frames. The motion generated from the single frame and multiple frame upsampling techniques may appear convincing and improve perceived video quality. However, the primary uses of high-speed cameras are scientific in nature, and interpolated/upsampled frames cannot be used to detect/track/classify high-speed phenomena.

\textbf{Event camera} offers an alternative pixel architecture with in-hardware compression designed for moving scene data~\cite{gallego2019event}. The pixel sensor readout circuit is designed to detect and report intensity changes in real-time, sparsifying sensor data by dynamically ignoring areas not detecting change. Removing the need for large and regular sensor readouts reduces data volume and latency. Unfortunately, event cameras also eliminate the majority of the static scene content needed to accurately reconstruct video. While methods do exist to reconstruct video from events~\cite{baldwin2021time,rebecq2019events}, these methods do not yet recreate low-contrast edges accurately due to the sensitivity limits inherent in today's event cameras.

%% file: 3_principle.tex
\begin{figure}[htbp]
\centering
\subfloat[Bayer Color Filter Array (CFA)]{\includegraphics[width=.95\linewidth]{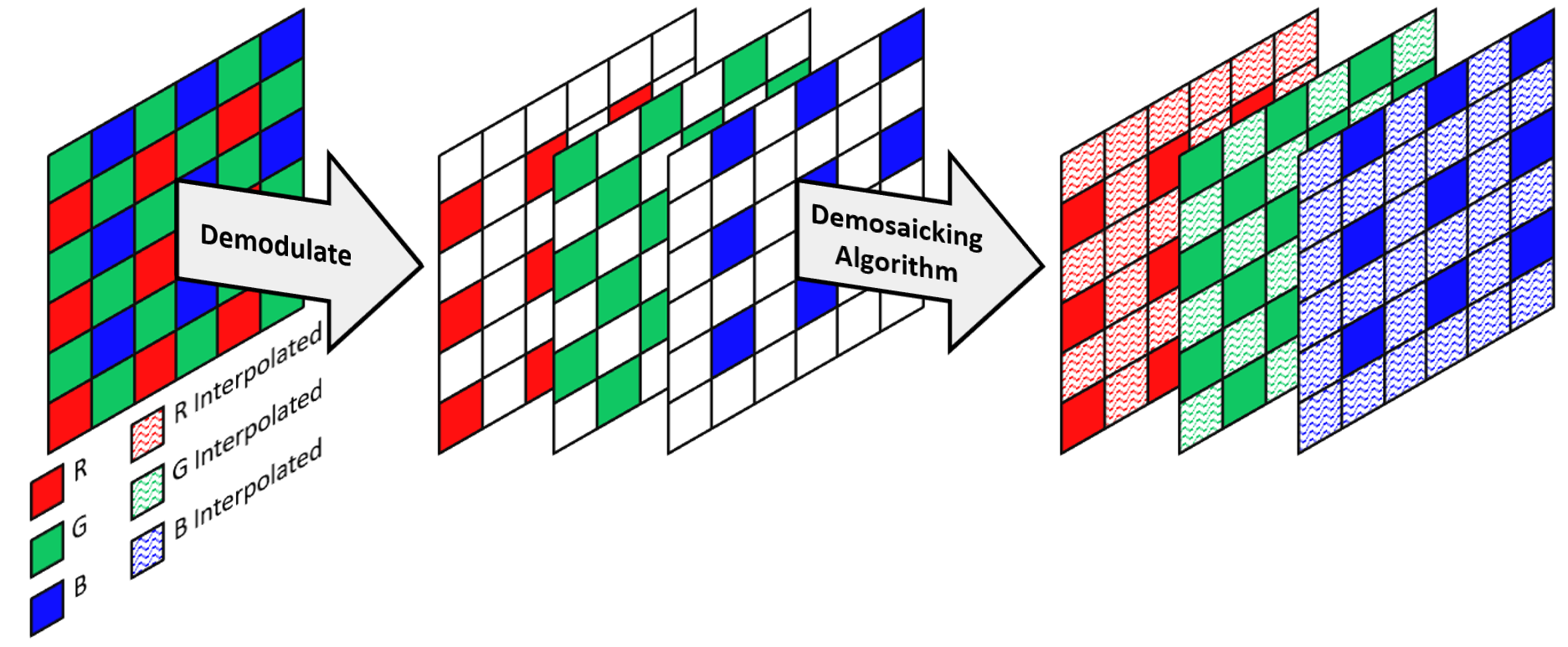}}\\[2ex]
\subfloat[Temporal Modulation Array (TMA)]{\includegraphics[width=.95\linewidth]{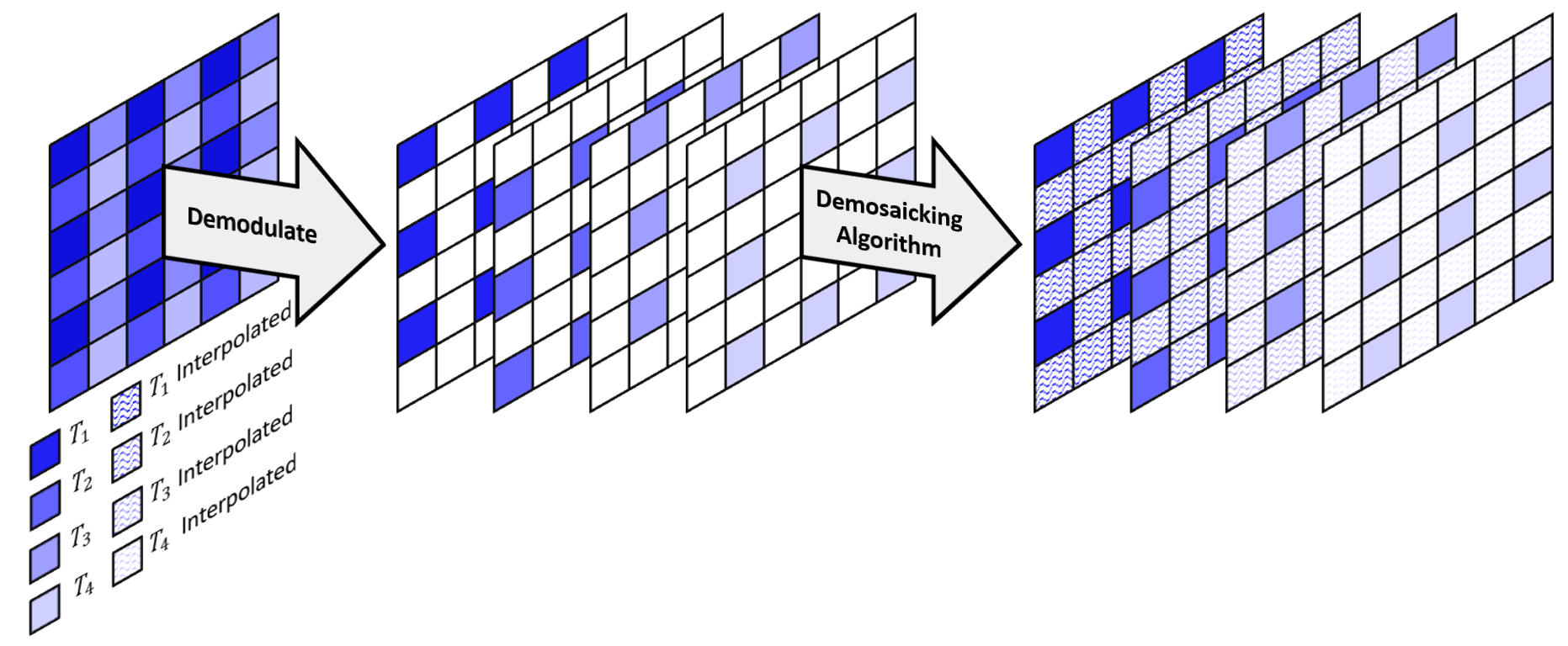}}
\caption{Bayer arrangement of color filter arrays of a typical imaging sensor. The sensor output is demosaicked to recreate a three-color image at full resolution. (left) Arrangement of temporal modulation arrays of a FC. The sensor output is demosaicked spatially and demodulated temporally to recreate an image sequence of four images at full resolution. (right)}
\label{fig:cfaVStfa}
\end{figure}

\section{Fourier Camera Design}
\label{sec:FC}

\subsection{Motivation: Hadamard Transform}
\label{sec:binaryHadamard}

There are three unique design features of the FC that differentiate it from existing coded exposure cameras and compressed sensing. First, our coded exposures use {\bf sine wave-like Walsh patterns} to encode temporal information. To enable this, our hardware configuration is designed to capture \textbf{sign-coded exposures}---unlike the conventional  \textbf{binary-coded exposures} that modulate light by partially blocking the light, the negative portion of the \textbf{sign-coded exposures} subtracts from the positive portion. We perform an inverse Hadamard transform to reconstruct the frames, which we show does not amplify noise. Second, FC spatially multiplexes Walsh functions over a pixel sensor array using a repeated pattern, similar to a color filter array (CFA) in a typical color camera. We refer to this repeating spatial multiplexed pattern as {\bf temporal modulation array (TMA)}, see Figure~\ref{fig:cfaVStfa}(b). The benefit to this design is that the \emph{demosaicking} method used to recover the complete Walsh function from its subsampled version is computationally efficient and noise-robust. Finally, appealing to the high degree of redundancies in high-speed video frames, TMA in FC is designed to sample temporal information less densely than the spatial signal. Such a sampling strategy represents an efficient TMA design that balances the spatial-temporal bandwidth of the high-speed image signal.


The key to extending a DMD camera framework to high-speed imaging is the condition number of the sensing matrix underlying the coded spatial-temporal exposure pattern---this is the main focus of our research. As already discussed in Section \ref{sec:priorwork}, the high condition number of pseudo-random binary sensing matrix used in compressive sensing techniques and the low-light efficiency of one-hot encoding limit the achievable frame rate of existing coded-pixel exposure systems. 

Towards the goal of extending the DMD camera framework to high-speed imaging operating at a very fast frame rate, a sensing matrix that (i) has a small condition number and (ii) has a large-light gathering property would be desirable. In this work, we consider Walsh functions used in the Hadamard transform, comprised of 1's and -1's and arranged in sine-wave like pattern to yield a unitary matrix (i.e.~condition number of 1). Contrasting to the aforementioned pseudo-random binary sensing matrix, the low condition number of the Walsh function patterns used by the proposed FC gives rise to a unitary transform known as Hadamard without risks of noise amplification or significant sacrifices to the photon count.




The Hadamard matrix is defined recursively as
\begin{align}
  \bm{H}_1 \stackrel{\triangle}{=} \left[ 
    \begin{array}{rr} 1 & 1 \\ 1 & -1 \end{array} \right ]\text{,}
\end{align}
\begin{align}
  \bm{H}_m=\bm{H}_1\otimes\bm{H}_{m-1}=\left[ \begin{array}{rr} 
      \bm{H}_{m-1} & \bm{H}_{m-1} \\ \bm{H}_{m-1} & -\bm{H}_{m-1} \end{array} \right] \text{,}
\end{align}
where $\otimes$ denotes the Kronecker product.
Let $f\in\mathbb{R}^{2^m}$ be a signal. Then the forward Hadamard transform can be written as
\begin{align}\label{eq:hadamard_coeff}
    \begin{bmatrix}h(0)\\h(1)\\\vdots\\h(2^m-1)\end{bmatrix}=\bm{H}_m\begin{bmatrix}f(0)\\f(1)\\\vdots\\f(2^m-1)\end{bmatrix},
\end{align}
and indeed the condition number of the matrix $\bm{H}_m$ is 1. Alternatively, the forward Hadamard transform $h(u)$ of the function $f(x)$ can be written as an inner-product with Walsh function, expressed as a series of sums and differences as follows:
\begin{align}
h(u)=\sum_{x=0}^{2^m-1}f(x)(-1)^{\sum_{i=0}^{m-1}b_i(x)b_{m-1-i}(u)}
\end{align}
where $b_i(x)$ is the $i$th bit of the length $m$ binary number $x$. Similarly, the inverse Hadamard transform can be computed as
\begin{align}
\label{eqn:inverseHad}
f(x)=\frac{1}{2^m}\sum_{u=0}^{2^m-1}h(u)(-1)^{\sum_{i=0}^{m-1}b_i(x)b_{m-1-i}(u)}.
\end{align}

The Walsh function is also very close to the binary encoding pattern implementable by DMD cameras. Recall that the micromirror in DMD is a physical device that reflects or blocks light at each pixel. As such, subtraction (corresponding to -1 in Hadamard matrix $\bm{H}_m$) cannot be accounted for optically using the conventional DMD camera configuration aimed at partially blocking the light instead. In the next subsections, we offer two alternative DMD camera configurations that enable sign-coded exposures.

\subsection{Sign-Coded FC Design \#1}


Let us rewrite the Hadamard transform matrix as:
\begin{align}
\label{eq:hadamard}
\begin{split}
    \bm{H}_m&=\begin{bmatrix}
    1&1&1&1&\\
    1&-1&1&-1&\cdots\\
    1&1&-1&-1&\\
    1&-1&-1&1&\\
    &\vdots&&&\ddots
    \end{bmatrix}\\
    &=
    \begin{bmatrix}
    1&1&1&1&\\
    1&0&1&0&\cdots\\
    1&1&0&0&\\
    1&0&0&1&\\
    \vdots&&&&\ddots
    \end{bmatrix} + 
    \begin{bmatrix}
    0&0&0&0&\\
    0&-1&0&-1&\cdots\\
    0&0&-1&-1&\\
    0&-1&-1&0&\\
    &\vdots&&&\ddots
    \end{bmatrix}\\
    &=
    \underbrace{\begin{bmatrix}
    1&1&1&1&\\
    1&0&1&0&\cdots\\
    1&1&0&0&\\
    1&0&0&1&\\
    \vdots&&&&\ddots
    \end{bmatrix}}_{\bm{H}_{pos}} - 
    \underbrace{\begin{bmatrix}
    0&0&0&0&\\
    0&1&0&1&\cdots\\
    0&0&1&1&\\
    0&1&1&0&\\
    &\vdots&&&\ddots
    \end{bmatrix}.}_{\bm{H}_{neg}}
\end{split}
\end{align}
We refer to $\bm{H}_{pos}$ and $\bm{H}_{neg}$ as the ``positive'' and ``negative'' Hadamard matrices, respectively. Positive Hadamard shares the same pattern as the Hadamard matrix $\bm{H}_m$ except that -1 are respectively replaced by 0; Negative Hadamard replaces 1 with 0 and -1 with 1. With no negative values in the sensing matrices, a traditional coded exposure configuration (a sensor combined with a DMD) can capture the pattern from either matrices. 
A single DMD can be used to toggle the photons per pixel at the $\bm{H}_{pos}$ or $\bm{H}_{neg}$ sensor. We refer to this configuration as \textbf{Sign-Coded FC Design \#1} and is shown in Figure~\ref{fig:camera_design}(b). The \emph{positive coded camera} and the \emph{negative coded camera} capture the following, respectively:
\begin{align}\label{eq:ac_posneg}
AC_{pos}(u) = \sum_{x=0}^{2^m-1} \frac{f(x)}{2} \left(1+(-1)^{\sum_{i=0}^{m-1}b_i(x)b_{m-1-i}(u)}\right)\\
AC_{neg}(u) = \sum_{x=0}^{2^m-1} \frac{f(x)}{2} \left(1-(-1)^{\sum_{i=0}^{m-1}b_i(x)b_{m-1-i}(u)}\right).
\end{align}
The Hadamard DC coefficient $h(0)$ may be computed by the post-capture summation:
\begin{align}\label{eq:d1_dc_posneg}
    h(0)=\sum_{x=0}^{2^m-1}f(x)=AC_{pos}(u) + AC_{neg}(u)
\end{align}
On the other hand, the post-capture difference between the 
the measured positive and negative coefficients yields the AC Hadamard coefficients in \eqref{eq:hadamard_coeff}:
\begin{align}\label{eq:d1_ac_posneg}
    h(u)=AC_{pos}(u) - AC_{neg}(u).
\end{align}
Applying inverse Hadamard to $h(0),\dots,h(2^m-1)$ via \eqref{eqn:inverseHad} recovers the high speed frame signal $f(0),\dots,f(2^m-1)$. 

Besides the remarkably low condition number of 1, Design \#1 has the advantage of 100\% light efficiency, as shown in Figure \ref{fig:TimeSignals}(c). In practice, however, Design \#1 is challenging to construct because typical DMD devices angle light by only about $\pm 12$ degrees. The tight clearance between the objective and macro lenses makes it difficult to fit all the optical and sensing components. This configuration is also incompatible with the total internal reflection (TIR) prisms that allow the DMD to be perpendicular to the optical axis (to make focusing at the detector easier).


\subsection{Sign-Coded FC Design \#2}
\label{sec:design2}


\textbf{Sign-Coded FC Design \#2}, which we develop below, is more practical. Suppose we rewrite the negative Hadamard matrix $\bm{H}_{neg}$ as:
\begin{align}
\label{eq:neghad}
\begin{split}
    \bm{H}_{neg}&=
\underbrace{
    \begin{bmatrix}
    1&1&1&1&\\
    1&1&1&1&\cdots\\
    1&1&1&1&\\
    1&1&1&1&\\
    &\vdots&&&\ddots
    \end{bmatrix}}_{\bm{H}_{dc}} - 
\underbrace{
\begin{bmatrix}
    1&1&1&1&\\
    1&0&1&0&\cdots\\
    1&1&0&0&\\
    1&0&0&1&\\
    &\vdots&&&\ddots
    \end{bmatrix}}_{\bm{H}_{pos}}.
\end{split}
\end{align}
Substituting this into \eqref{eq:hadamard}, we now have $\bm{H}_m$ written in terms of $\bm{H}_{pos}$ and $\bm{H}_{dc}$ as follows:
\begin{align}
\label{eq:poshad}
\begin{split}
    \bm{H}_m
    &=2\bm{H}_{pos}-\bm{H}_{dc}.
\end{split}
\end{align}
The matrix $\bm{H}_{dc}$ in \eqref{eq:neghad} and \eqref{eq:poshad} is simply the DC component of the light signal and is exactly the same signal that would be captured by a typical camera with no DMD device (i.e. ``always on''). This \textbf{Sign-Coded FC Design \#2} can be physically implemented by a two-camera configuration shown in Figure~\ref{fig:camera_design}(c). A beamsplitter is used to split the light to one camera that only records a DC value at every pixel (``DC camera''), and a second camera with TMA spatial multiplexing to encode and capture positive Hadamard components (``AC camera''). The image captured by the DC camera is neither spatially nor temporally modulated:
\begin{align}
DC = \sum_{x=0}^{2^m-1} f(x) = h(0).
\end{align}
On the other hand, the image captured by the AC camera is spatial-temporally modulated by the DMD using the positive Hadamard code, and reflected by TIR prism:
\begin{align}\label{eq:ac}
AC(u) = \sum_{x=0}^{2^m-1} \frac{f(x)}{2} \left(1+(-1)^{\sum_{i=0}^{m-1}b_i(x)b_{m-1-i}(u)}\right)
\end{align}
The Hadamard coefficients in \eqref{eq:hadamard_coeff} are reconstructed from the measured DC and AC coefficients in post-capture processing:
\begin{align}\label{eq:twocamera}
    h(u)=2\cdot AC(u) - DC.
\end{align}
Applying inverse Hadamard to $h(0),\dots,h(2^m-1)$ via \eqref{eqn:inverseHad} recovers the high speed frame signal $f(0),\dots,f(2^m-1)$. 

\begin{figure*}
    \centering
    \begin{tabular}{@{}c@{~}c@{~}c@{~}c@{~}c@{}}
       \includegraphics[width=.19\textwidth]{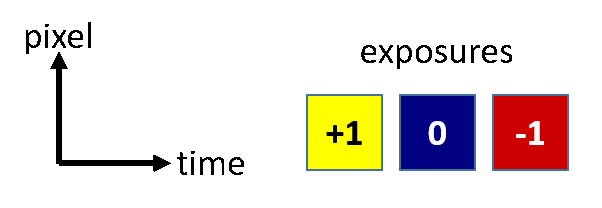}& \includegraphics[width=.19\textwidth]{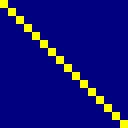}&      \includegraphics[width=.19\textwidth]{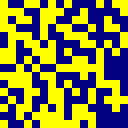}&
        \includegraphics[width=.19\textwidth]{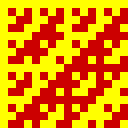}&
        \includegraphics[width=.19\textwidth]{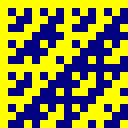}\\
&
          (a) One-Hot\cite{bub2010temporal}&
          (b) Pseudo-Random &
          (c) Signed Hadamard&
          (d) Positive Hadamard\\
    \end{tabular}
        \vspace{-8pt}
    \caption{Length-16 coded exposures. (a) One-hot encoding activates exposure for a brief period of time, blocking most of the light. (b) Binary-coded pseudo-random exposure has 50\% light efficiency. (c) Sign-coded Hadamard is used by the proposed sign-coded FC Design \#1 and has 100\% light efficiency, where the negative portion of the exposure is captured by a second camera. (d) The positive Hadamard is also a binary-coded pattern, used as a part of the proposed sign-coded FC Design \#2. It can be used in conjunction with a second DC camera to compute a signed Hadamard exposure pattern in post-processing.}
    \label{fig:TimeSignals}
\end{figure*}

Like binary-coded pseudo-random exposure pattern, the timing diagram in Figure~\ref{fig:TimeSignals}(d) shows that far more light would reach the sensor in AC camera using the positive Hadamard sensing when compared to the one-hot approach in Figure~\ref{fig:TimeSignals}(a) (though not as much as FC Design \#1). Yet, as shown by Table \ref{tab:cond_number}, FC Design \#2 has a fixed condition number that is orders of magnitude lower than the binary-coded pseudo-random exposure pattern. For this reason, FC Design \#2 yields higher SNR reconstruction compared to the binary random sensing matrix. 

Comparing the two proposed FC configurations, Design \#1's Hadamard coefficient reconstruction step in \eqref{eq:d1_dc_posneg} and \eqref{eq:d1_ac_posneg} are orthogonal, whereas Design \#2's equivalent step in \eqref{eq:twocamera} is not, accounting for the higher condition number of Designs \#2. In practice, Design \#2 is vulnerable to calibration errors in zero offset, which is a constant offset in every DC and AC measurement. Rewriting \eqref{eq:twocamera} with offset $\eta$, we have:
\begin{align}\label{fig:withoffset}
    2\cdot (AC(u)+\eta)-(DC+\eta)=h(u)+\eta.
\end{align}
That is, the offset $\nu$ remains in each Hadamard coefficient. By linearity, the inverse Hadamard transform of the constant offset $\eta$ would be added to the reconstructed high-speed frames. Rewriting \eqref{eqn:inverseHad} to include this offset, we have
\begin{align}
\begin{split}
&\frac{1}{2^m}\sum_{u=0}^{2^m-1}(h(u)+\eta)(-1)^{\sum_{i=0}^{m-1}b_i(x)b_{m-1-i}(u)}=f(x)+\eta\delta(x),
\end{split}
\end{align}
meaning the effects of offset error is confined to the frame time $x=0$ only. Proper calibration to calibrate out $\eta$ from the raw sensor data would further improve the quality of 0th frame reconstruction. The Design \#1 on the other hand is less susceptible to offset errors since $\eta$ is canceled in \eqref{eq:d1_ac_posneg}.



As a side note, one can design another stable coded-exposure system using only positive Hadamard matrix $\bm{H}_{pos}$ defined in \eqref{eq:hadamard} (i.e.~use only one AC camera, without the DC camera). This ``binary-coded Fourier Camera'' pattern can be implemented on a simpler one-camera hardware configuration in Figure \ref{fig:camera_design}(a). Though the condition number is higher than the sign-coded FC, it is still far lower than the binary-coded pseudo-random exposure pattern (see Table \ref{tab:cond_number}) and higher light efficiency than one-hot encoding (see Figure \ref{fig:TimeSignals}(a)).


\section{Spatial Light Modulation Design}
\label{sec:tfa}



\begin{figure}
    \centering
            \includegraphics[width=.48\textwidth]{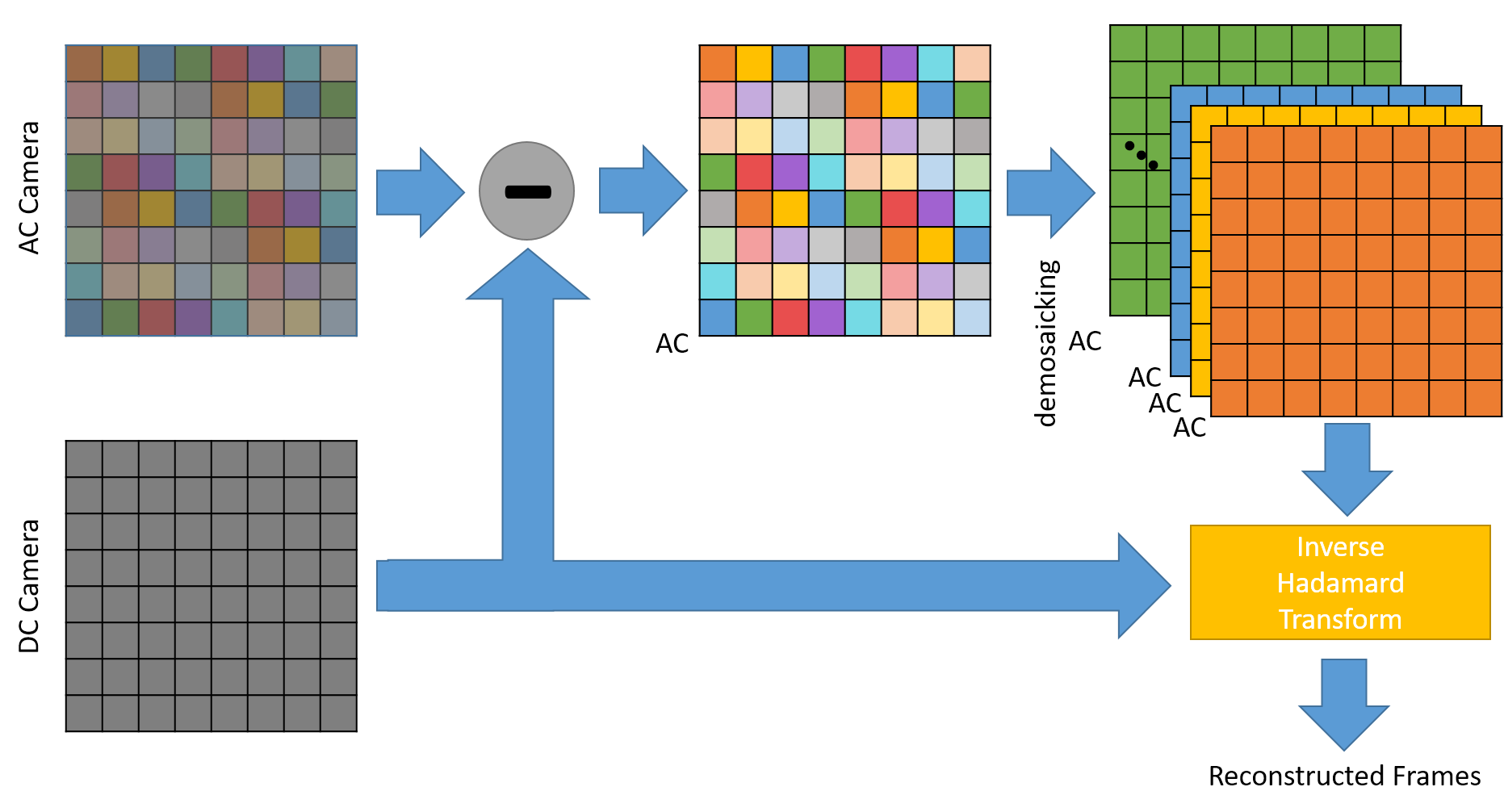} 
    \caption{Frame reconstruction for Sign-Coded FC Design \#2 via demosaicking and inverse Hadamard transform. Hadamard's DC coefficient is observed at every pixel, which is subtracted from positive Hadamard coefficients to yield signed Hadamard values. See Section \ref{sec:design2}}
    \label{fig:demosaicking}
\end{figure}




Recall the \textbf{Sign-Coded FC Design \#2} shown in Figure~\ref{fig:camera_design}(c). The DC camera yields a dense, high-resolution baseline image. This is in contrast to the AC camera, which spatially multiplexes the temporally modulated coded-exposure patterns as shown by Figure \ref{fig:TimeSignals}(d). Unlike the binary-coded pseudo-random exposure patterns employed by compressive sensing approaches, we propose a lattice pattern that draws on color filter array (CFA) designs in camera image sensors (shown in Figure~\ref{fig:cfaVStfa}(a)). Specifically, CFA is a spatial multiplexing of color filters that sacrifices spatial resolution for additional spectral measurements. One major benefit to this approach is that the reconstruction via an interpolation process referred to as \emph{demosaicking} is computationally efficient. Like CFA, the proposed FC design employs spatial multiplexing of exposure codes we refer to as temporal modulation array (TMA) to trade spatial resolution for additional temporal resolution (see Figure~\ref{fig:cfaVStfa}(b)). The corresponding high-speed frame reconstruction strategy is outlined in Figure \ref{fig:demosaicking}, which we detail below. 

\begin{figure*}
    \centering
\begin{tabular}{@{}c@{~}c@{~}c@{~}c@{~}c@{~}c@{}}
    \includegraphics[width=0.155\textwidth]{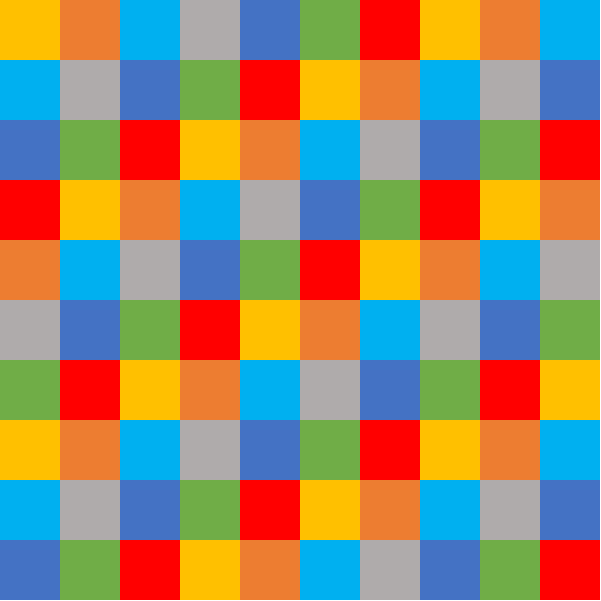}&
    \includegraphics[width=0.155\textwidth]{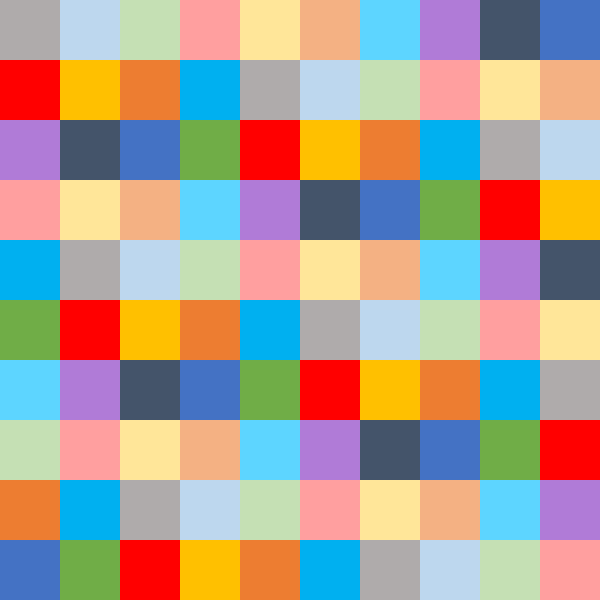}&
    \includegraphics[width=0.155\textwidth]{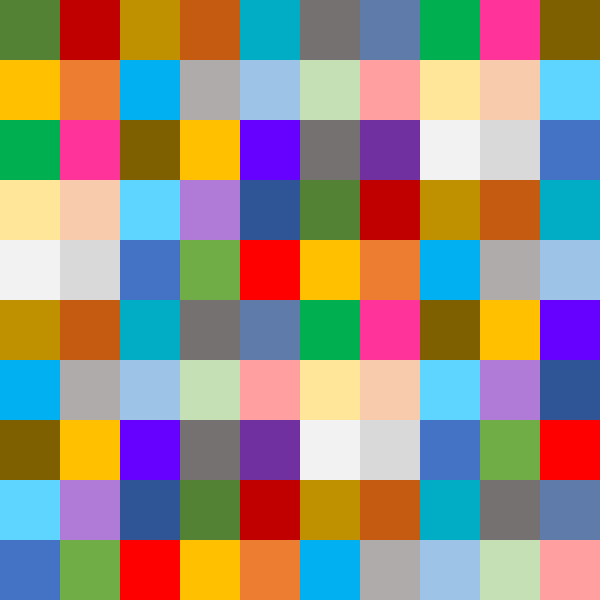}&
    \includegraphics[width=0.165\textwidth]{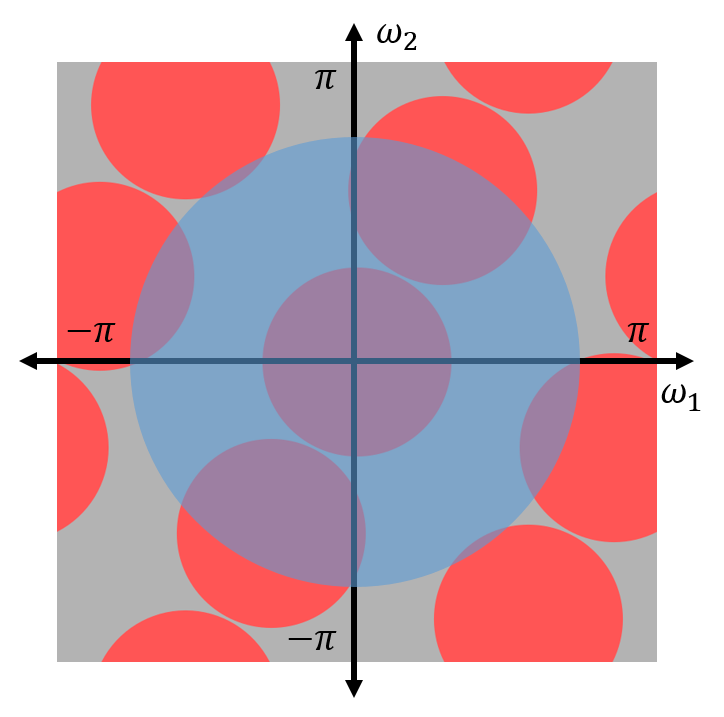}&
    \includegraphics[width=0.165\textwidth]{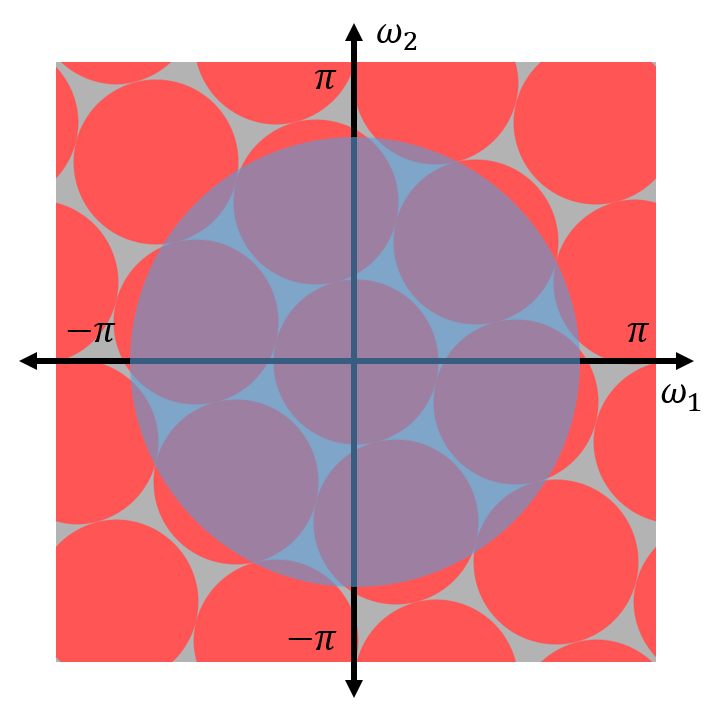}&
    \includegraphics[width=0.165\textwidth]{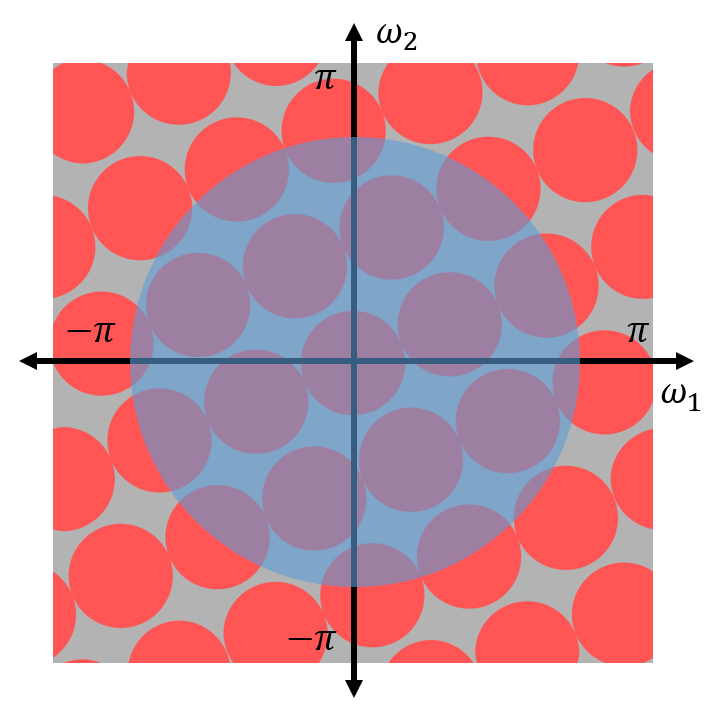}\\
    (a) 7 AC TMA &
    (b) 15 AC TMA &
    (c) 31 AC TMA &
    (d) FT~of (a)&
    (e) FT~of (b)&
    (f) FT~of (c)
\end{tabular}
    \caption{(a-c) Spatial light modulator pattern (TMA) designs used in AC camera, organized as integer hexagonal lattices generated by $M_3$, $M_4$, and $M_5$ in \eqref{eq:slm}.  (d-f) Corresponding Fourier transform of SLM image in \eqref{eq:modulation}. Blue=$H(\bf{\omega},0)$ that can be subtracted out by DC camera. Red=linear combinations of TMA-modulated Hadamard coefficients $\{H(\bm{\omega},1),\dots,H(\bm{\omega},2^m-1)\}$.}
    \label{fig:fft}
\end{figure*}

The spatial arrangement of SLM using the DMD determines the spatial bandwidth supported by the sensor. Drawing on the CFA pattern design in \cite{hirakawa2008spatio}, we develop a novel ``integer hexagonal'' lattice-based TMA pattern to maximize the spatial bandwidth by minimizing the risk of aliasing. Suppose the spatial sampling stemming from TMA falls on the following lattice:
\begin{align}\label{eq:lattice}
    \bm{\Lambda}:=\bm{M}\mathbb{Z}^2=\{\bm{k}\in\mathbb{Z}^2|\bm{k}=\bm{M}\bm{q},\bm{q}\in\mathbb{Z}^2\},
\end{align}
where the integer ``generator'' matrix $\bm{M}\in\mathbb{Z}^{2\times 2}$ determines the lattice pattern. This lattice pattern has exactly $|\det (\bm{M})|$ cosets (non-overlapping shifted versions of the lattice):
\begin{align}\label{eq:coset}
    \bm{\ell}+\bm{\Lambda},\qquad \forall \bm{\ell}\in \bm{M}[0,1)^2\cap \mathbb{Z}^2.
\end{align}
Some examples are shown in Figure~\ref{fig:fft}(a-c).

Regarding cosets as spatial multiplexing (assign a specific AC component $AC(\bm{k},u)$ to each coset $\bm{\ell}+\bm{\Lambda}$), a spatial light modulator with this lattice-based TMA pattern can support up to $N=|\det (\bm{M})|$ distinct coded exposure patterns. That is, the SLM image captured by the proposed FC with this TMA pattern is
\begin{align}\label{eq:slm}
\begin{split}
    g(\bm{k})=&\sum_{u=1}^{N-1} \sum_{\bm{\lambda}\in\bm{M}\mathbb{Z}^2} AC(\bm{k},u)\delta(\bm{k}-\bm{\lambda}-\bm{\ell}_u)\\
    =&\frac{h(\bm{k},0)}{2}+\sum_{u=1}^{N-1} \sum_{\bm{\lambda}\in\bm{M}\mathbb{Z}^2}  \frac{h(\bm{k},u)}{2}\delta(\bm{k}-\bm{\lambda}-\bm{\ell}_u),
\end{split}
\end{align}
where $\{\bm{\ell}_0,\bm{\ell}_1,\dots,\bm{\ell}_{N-1}\}$ refer to N distinct cosets in $\bm{M}[0,1)^2\cap \mathbb{Z}^2$. Recalling Figure \ref{fig:demosaicking}, subtracting out the DC camera image cancels $h(\bm{k},0)=DC(\bm{k})$:
\begin{align}\label{eq:ac_dc}
    2g(\bm{k})-DC(\bm{k})=\sum_{u=1}^{N-1} \sum_{\bm{\lambda}\in\bm{M}\mathbb{Z}^2}  h(\bm{k},u)\delta(\bm{k}-\bm{\lambda}-\bm{\ell}_u).
\end{align}
The Fourier transform of the subtraction residual in \eqref{eq:ac_dc} is:
\begin{align}\label{eq:modulation}
    \sum_{\bm{\nu}\in 2\pi\bm{M}^{-T}\mathbb{Z}^2\cap[-\pi,\pi)^2}\sum_{u=1}^{N-1}H(\bm{\omega}-\bm{\nu},u) \frac{e^{-j\bm{\nu}^T\bm{\ell}_u}}{N} ,
\end{align}
where $\bm{\omega}\in[-\pi,\pi)^2$ is spatial frequency; and $H(\bm{\omega},u)$ denote discrete space Fourier transforms of the Hadamard coefficients $h(\bm{k},u)$, respectively. Here, $\bm{\nu}\in[-\pi,\pi)^2$ is a spatial modulation frequencies (shifting of the spatial frequency by $\bm{\nu}$) stemming from subsampling in TMA, 
and $e^{-j\bm{\omega}^T\bm{\ell}_u}$ is the phase term induced by the coset lattice translation. See Figure \ref{fig:fft}(d-f).

The choice of integer generator matrix $\bm{M}\in\mathbb{Z}^{2\times 2}$ completely determines the spatial modulation frequencies. Aliasing occurs when the support of modulated signal
$H(\bm{\omega}-\bm{\nu},u)$ overlaps another modulated component $H(\bm{\omega}-\bm{\nu}',u')$.
Thus, maximizing distance  $\|\bm{\nu}-\bm{\nu}'\|_2$ in a lattice structure $\bm{\nu}\in 2\pi\bm{M}^{-T}\mathbb{Z}^2$ reduces aliasing risks. This is known as the ``sphere packing'' problem, whose solution in two dimensions is widely known to be a hexagonal lattice. In our work, we used an integer hexagonal lattice pattern that approximates hexagonal, using 
the generator matrix $\bm{M}\in\mathbb{Z}^{2\times 2}$ are as follows:
\begin{align}\label{eq:generator1}
    \bm{M}_3=\begin{bmatrix}
    2&3\\
    1&-2
    \end{bmatrix},\quad
    \bm{M}_4=\begin{bmatrix}
    3&4\\
    3&-1
    \end{bmatrix},\quad
    \bm{M}_5=\begin{bmatrix}
    2&7\\
    5&2
    \end{bmatrix}.
\end{align}
By exhaustive search, the above matrices were found to best approximate hexagonal lattice among all integer combinations yielding $N=|\det(\bm{M})|=2^m-1$ for $m=3,4,5$, respectively (since there are a total of $2^m-1$ AC components in \eqref{eq:ac}). As shown in Figure~\ref{fig:fft}(d-f), the resultant tessellations of Fourier coefficients stemming from the integer hexagonal TMA patterns are also approximately hexagonal, thereby maximizing the spatial bandwidth of the FC.

Continuing to follow the steps in Figure~\ref{fig:demosaicking}, the purpose of demosaicking is to reconstruct full resolution $h(\bm{k},u)$ from their TMA sampled version in \eqref{eq:ac_dc}. Though the choices of demosaicking are empirically explored in Section \ref{sec:simResults} below, we briefly describe a baseline method called frequency selection \cite{dubois2005frequency}. Specifically, demodulation is carried out by applying a lowpass filter $\phi(\bm{k})$ to the product of the modulated signal $2\cdot g(\bm{k})-h(\bm{k},0)$ and the carrier $e^{j\bm{\nu}^T\bm{k}}$:
\begin{align}
    \phi(\bm{k})\star \{(2\cdot g(\bm{k})-h(\bm{k},0))e^{j\bm{\nu}^T\bm{k}}\}
=\sum_{u=1}^{N-1}h(\bm{k},u)e^{-j\bm{\nu}^T\bm{\ell}_u}.
\end{align}
Repeating this procedure for all modulation components $\{\bm{\nu}_1,\dots,\bm{\nu}_{N-1}\}\in 2\pi\bm{M}^{-T}\mathbb{Z}^2$ yields the relation:
\begin{align}
\begin{split}
&\phi(\bm{k})\star\left\{(2\cdot g(\bm{k})-h(\bm{k},0))    
\begin{bmatrix}
   e^{j\bm{\nu}_1^T\bm{k}}\\
   \vdots\\
   e^{j\bm{\nu}_{N-1}^T\bm{k}}
    \end{bmatrix}
\right\}\\
    &=
    \underbrace{\begin{bmatrix}
    e^{-j\bm{\nu}_1^T\bm{\ell}_1}&\cdots&e^{-j\bm{\nu}_1^T\bm{\ell}_{N-1}}\\
    \vdots&\ddots&\vdots\\
    e^{-j\bm{\nu}_{N-1}^T\bm{\ell}_1}&\cdots&e^{-j\bm{\nu}_{N-1}^T\bm{\ell}_{N-1}}\\
    \end{bmatrix}}_{\bm{E}\in\mathbb{R}^{N-1\times N-1}}
    \begin{bmatrix}
    h(\bm{k},1)\\\vdots\\h(\bm{k},N-1)
    \end{bmatrix}.
\end{split}
\end{align}
Applying inverse matrix $\bm{E}^{-1}$ recovers Hadamard coefficients $\{h(\bm{k},1),\dots,h(\bm{k},N-1)\}$. Inverting the Hadamard matrix $\bm{H}_m$ in \eqref{eq:hadamard_coeff} reconstructs the full-resolution high-speed frames.

%% file: 4_design.tex
\section{Prototype Design}


\subsection{Hardware Configuration}

We prototyped \textbf{Sign-Coded FC Design \#2} using the DMD from Texas Instruments LightCrafter\textsuperscript{TM} 4500 evaluation module~\cite{lightcrafter20154500}. The LightCrafter is composed of both a light engine and a driver board. The light engine contains optics, LEDs, and a 912$\times$1140 diamond pixel 0.45-inch WXGA DMD. The driver board contains flash memory, a driver circuit, a DMD controller, and I/O ports. A DMD acts as a Spatial Light Modulator (SLM) to steer visible light and create adjustable binary patterns at very high frame rates. The TI DLP4500 works at frame rates up to 4.5kHz, but newer models, such as the TI DLP7000 and DLP9000, work at higher resolutions and allow for binary patterns up to 32kHz. The LightCrafter was disassembled and mounted to an optics bench to gain direct access to the DMD.

\begin{figure}[htbp]
    \centerline{\includegraphics[width=1\linewidth]{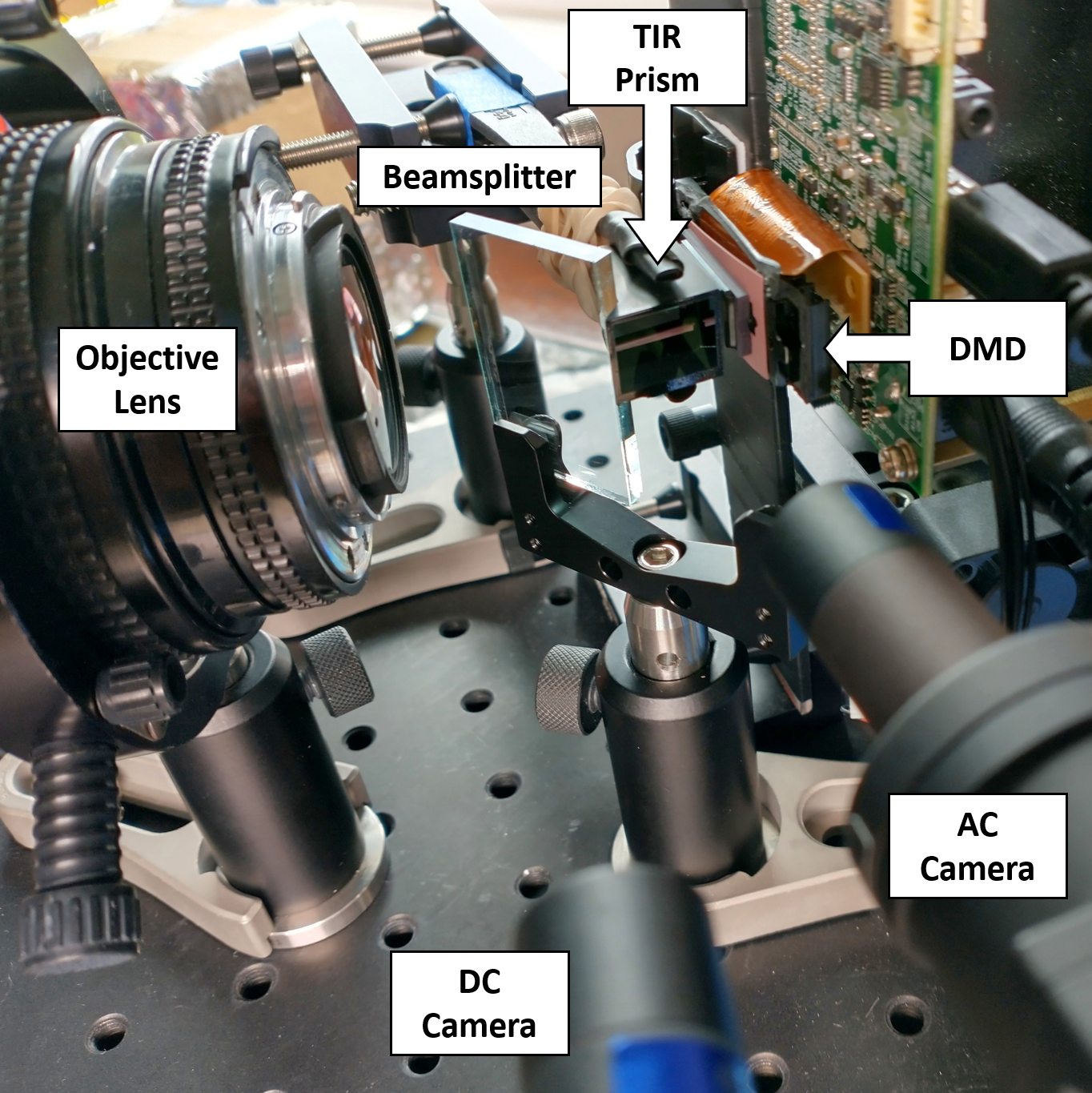}}
    \caption{Assembled Sign-Coded FC including objective lens, TIR prism, DMD, and two FLIR cameras. Light enters through the Nikon DSLR lens and refracts through the TIR prism before reaching the DMD. The DMD encodes the light by passing or blocking light per micromirror. Passed light reflects off the prism and is focused on the imaging sensor via a macro lens.}
    \label{fig:cameraClose}
\end{figure}

Figure~\ref{fig:camera_design}(c) diagrams the schematic of the FC optical path, and its physical implementation is shown in Figure~\ref{fig:cameraClose}. Light is focused using an objective lens (i.e. Nikon 28-85mm AF DSLR lens). Like many other DSLR lenses, this lens is near-telecentric on the image side to eliminate color cross-talk. It is beneficial in our application to help decrease the depth-of-focus since the DMD is not perfectly perpendicular to the optical axis. The imaging sensors are a matched pair of FLIR Blackfly\textregistered~S USB3. Each sensor is a 1/1.8'' format 3.2MP monochrome camera (2048$\times$1536 resolution) capable of framing at 118 FPS. Its global shutter helps avoid complex timing issues with the DMD. The AC camera is equipped with an Opto Engineering MC075X macro lens. This macro lens has a working distance of 58mm with a 9.5 $\times$ 7.2mm field-of-view (DMD is 9.855 $\times$ 6.161mm). The DC camera has an Opto Engineering MC050X macro lens. 


\begin{figure}[htbp]
    \centerline{\includegraphics[width=0.95\linewidth]{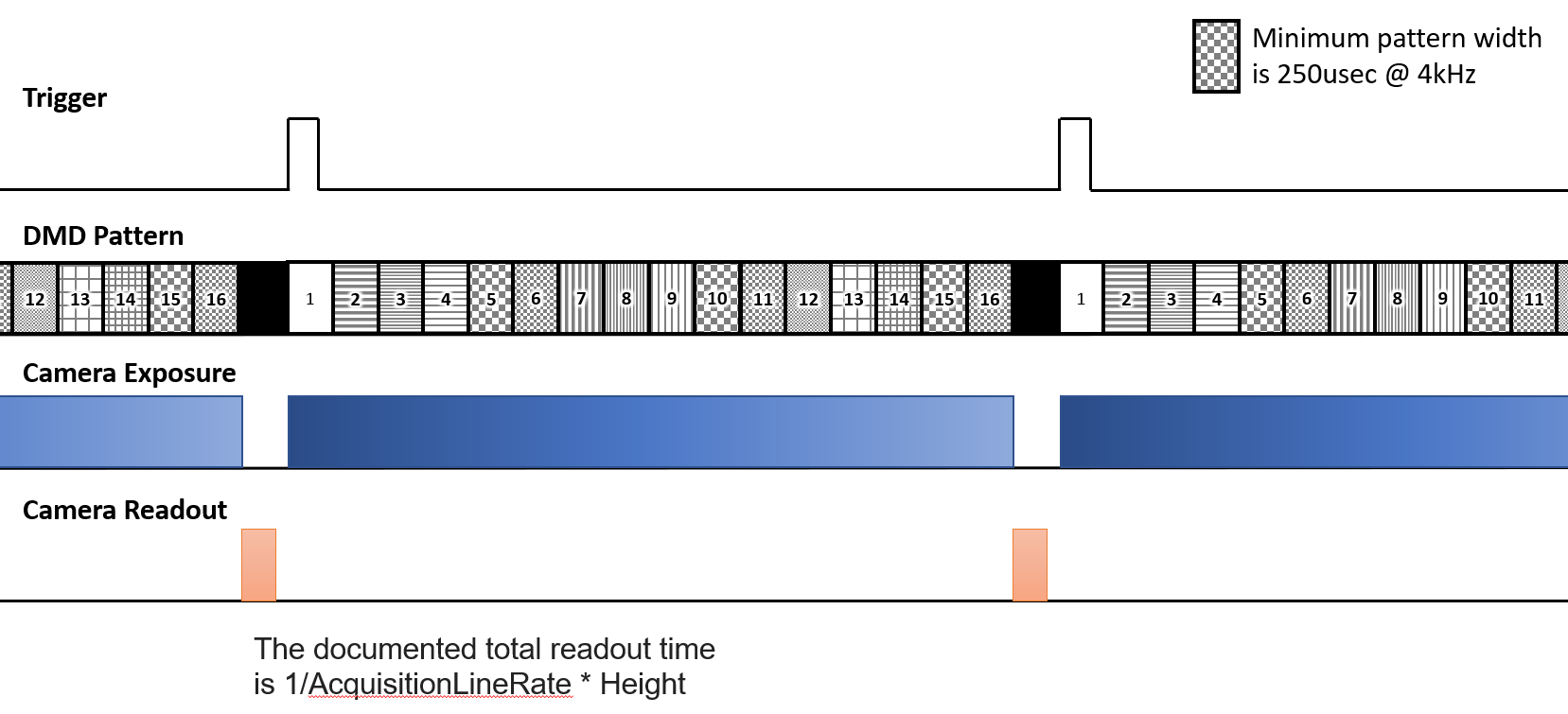}}
    \caption{A rising-edge trigger signal is used to synchronize the DMD patterns and the camera exposure. At the rising edge of the trigger signal, the DMD starts a 16-pattern sequence and at the same time the camera exposure begins. The DMD runs through all 16 patterns during the single exposure. At the end of the final pattern, the camera exposure ends and the image readout begins. The system is configured to allow sufficient readout time before the next trigger signal.}
    \label{fig:timingDiagram}
\end{figure}

Precise time synchronization is required between the DMD and the imaging sensors to achieve maximum frame rates and minimal noise. We used the LightCrafter to trigger FLIR cameras connected via a Hirose HR10 (6-pin) GPIO cable. As shown by the timing diagram in Figure~\ref{fig:timingDiagram}, the rising edge of the trigger signal is both the start of the first coded pattern and the start of exposure on the camera. The delay from trigger to pattern exposure and the start of image capture is less than \SI{1}{\us}. DMD mirrors transition states in less than \SI{5}{\us}.



%% file: 5_challenges.tex
\subsection{Calibration and Crosstalk}
\label{sub:tfa}

\begin{figure}
    \centering
    \begin{tabular}{c@{~}c@{~}}
        \includegraphics[width=.23\textwidth]{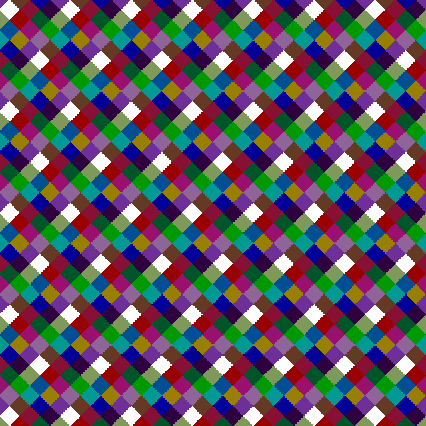}&
        \includegraphics[width=.23\textwidth]{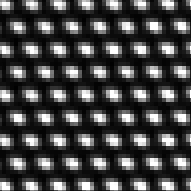}\\
          (a) TMA Pattern & (b) Raw Sensor Data 
    \end{tabular}
    \caption{The DMD micromirror in the proposed Binary Coded FC prototype are arranged in quincunx grid. 
    (a) Length-16 coded exposure on DMD quincunx micromirror grid. To minimize the risk of aliasing, the proposed TMA is based on an integer hexagonal lattice. 
    (b) One TMA pattern activated on DMD (i.e. only the mirrors represented in white are on), observed by the AC camera. A single micromirror does not converge to a point because the lens cannot resolve the detail. If the point spread function is broad, there is a high risk of crosstalk.}
    \label{fig:spatialCrosstalkTest1x1}
\end{figure}

Due to the different grid sizes and shapes between the DMD and the two FLIR cameras, the TMA pattern on DMD does not have a one-to-one mapping to camera pixels. To calibrate between DMD mirrors and AC camera pixels, the objective lens of the FC is defocused while it is pointed at a uniform white surface. DMD forms 44 graycode patterns that are recorded by the AC camera, yielding a homographic mapping between DMD mirrors and AC camera pixels. If necessary, we adjust the distance between the TIR prism and the AC camera to ensure that the graycode pattern displayed on DMD appears in focus to the AC camera. Note that the DMD mirrors are diagonally oriented (i.e.~organized in quincunx lattice), as shown in Figure \ref{fig:spatialCrosstalkTest1x1}(a)---this is not a problem as the homographic mapping can appropriately capture the rotation between DMD and AC camera.

Next, we calibrate between the AC and DC cameras as follows. We activate all DMD mirrors such that no light is blocked from AC camera; the objective lens is focused on a checkerboard calibration target such that it appears sharp in the AC camera (which is also sharp on DMD). We then adjust the distance between the DC camera and the beamsplitter so that the same calibration target appears sharp in the DC camera. Once the focus is set, homography mapping between the AC and DC camera pixels is computed from the corner points detected from the checkerboard pattern. In practice, we varied the positions of the calibration target throughout the scene to ensure robust calibration.

Figure \ref{fig:spatialCrosstalkTest1x1}(b) shows an image chip from an AC camera capturing one of the integer hexagonal lattice patterns in \eqref{eq:lattice} activated on the DMD mirrors. It is evident by Figure \ref{fig:spatialCrosstalkTest1x1}(b) that there is a risk of crosstalk between neighboring TMA samples due to the imaged mirror not converging to a point (i.e.~lens point spread function) on the AC camera. We address crosstalk in two ways. The most straightforward way is to downsample the DMD mirrors (either by a cuincunx or $2\times 2$ square lattice) such that the TMA samples are spaced farther away from each other. Although this comes at the cost of spatial resolution loss, the risks of crosstalk is drastically reduced by downsampling.

The second way to address crosstalk is by post-processing. In the prototype design, $912\times 1140$ micromirrors in DMD appear entirely within the field of view of the FLIR Blackfly camera covering $2048\times 1292$ pixels (approximately 2.5$\times$ oversampling). We model AC camera measurement as an over-determined system of linear equations
\begin{align}
\bm{A}\bm{u} =\bm{v}
\end{align}
where $\bm{u}$ is the vectorized intensity values at DMD mirrors (length $912\times 1140=1039680$); and $\bm{v}$ is the vectorized  pixel intensities recorded by the AC camera (length $2048\times 1292=2646016$). The pseudo-inverse of the matrix $\bm{A}$ (size $2646016 \times 1039680$) can reconstruct the intensities observed at each DMD mirror position. Owing to the fact that the matrix $\bm{A}$ is sparsely populated and highly over-determined, the pseudo-inverse is stable.

%% file: 6_experiments.tex
\section{Experiments}

\subsection{Feasibility of Temporal Demosaicking}
\label{sec:simResults}

\begin{figure*}
\begin{tabular}{@{}c@{}c@{}c@{}c@{}}
        \includegraphics[width=.25\textwidth]{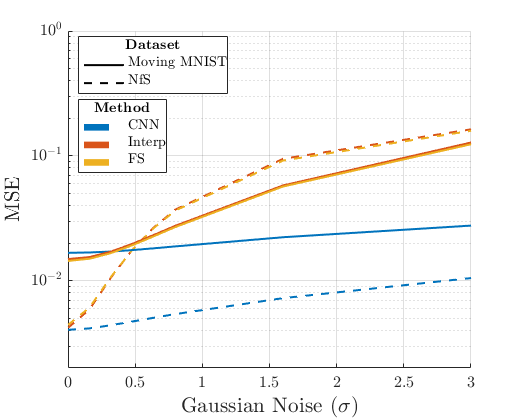} &  
         \includegraphics[width=.25\textwidth]{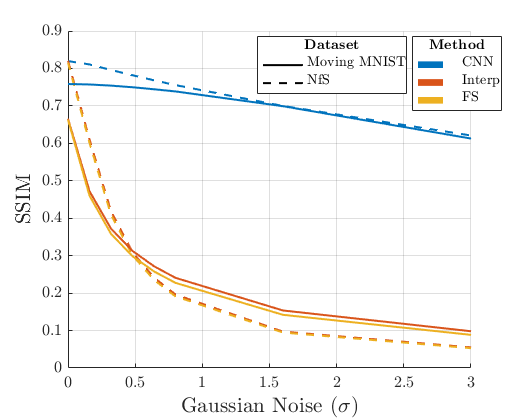}&
         \includegraphics[width=.25\textwidth]{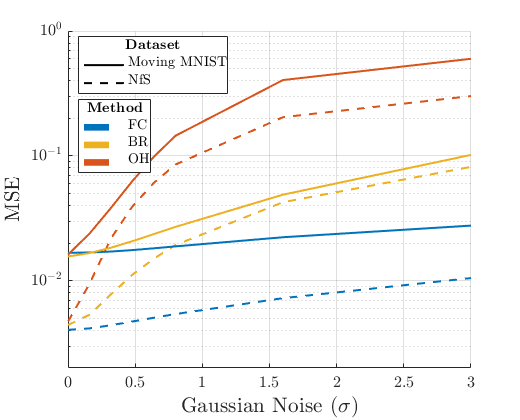} &  
         \includegraphics[width=.25\textwidth]{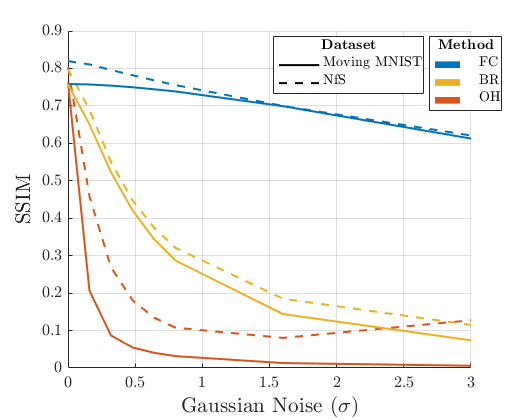}\\
         (a)&(b)&(c)&(d)
\end{tabular}
    \centering
   \caption{Evaluation of high-speed frame reconstruction qualities in terms of mean squared error (MSE) and structural similarity index measure (SSIM). Simulation performed on ``Moving MNIST''~\cite{srivastava2015unsupervised} and ``Need for Speed''~\cite{galoogahi2017need} datasets. Coded exposures were simulated by compressing 16 frames into a single camera image, with various levels of Gaussian noise added. (a,b) Comparisons of bilinear interpolation (Interp), frequency selection (FS), and U-Net demosaicking (CNN) \cite{ronneberger2015u} on the proposed Sign-Coded FC. Training data to CNN included noise so that it learns to handle noise, which unsurprisingly handled noise better than Interp and FS. (c,d) Comparisons of sign-coded FC, binary-coded one-hot (OH), and binary-coded random (BR) exposure patterns. Reconstruction was performed using the identical U-Net demosaicking architecture with the same training strategy. Overall, the proposed sign-coded FC was far more robust to noise than OH and BR coded exposure patterns.}
\label{fig:compare_coded_exposures}
\end{figure*}




We conducted a simulation study to explore different choices of TMA demosaicking methods for the proposed sign-coded FC to gauge feasibility, understand their output image quality, and investigate robustness to noise. Demosaicking has a rich history in research and the empirical evidence to restore visually pleasing full-resolution full-color images from the CFA-sampled sensor data. Demosaicking methods are usually implemented as single-pass process, unlike the time-consuming iterative linear programming methods commonly used in compressive sensing. Since temporal modulation demosaicking has limited prior work, we tested the feasibility and the performance of three \emph{representative} demosaicking strategies including bilinear interpolation, frequency selection (FS)~\cite{dubois2005frequency}, and a CNN. FS demosaicking method was already described in Section \ref{sec:tfa}.

For the CNN-based demosaicking method, we trained the network by simulating coded-exposure data from a high-speed video sequence. The full-resolution Hadamard transform coefficients, computed from the original video sequence via \eqref{eq:hadamard_coeff}, were used as the ground truth data. To explore noise robustness, we also corrupted the training coded exposure data with additive Gaussian noise at various noise levels. We used the same U-Net architecture~\cite{ronneberger2015u} across all tests with an encoder-decoder depth of 2 and the number of filters set to 16, 32, and 64 for each layer. The U-Net was attached to a $1\times1$ convolutional layer to match the desired number of output frames before the final MSE regression layer. The $3\times3$ convolutions were zero-padded so feature size did not change and utilized He initialization\cite{he2015delving}.

Testing was done on two different datasets. First, ``Moving MNIST''~\cite{srivastava2015unsupervised} consists of 10,000 sequences of 20 frames each. The images are $64\times64$ each and show 2 digits moving. Although simulated video, this dataset is challenging due to the high contrast and non-linear motion. We also tested using the ``Need for Speed'' (NfS) dataset~\cite{galoogahi2017need}. This dataset is constructed from 100 real-world scenarios with videos at 240FPS and includes challenging scenes with occlusion, fast motion, noise, and clutter. 

For both datasets, we extracted cropped samples (``chips'') of size 64$\times$64 pixels and 16 frames. We used 150 randomly selected chips per video sequence, and avoided temporally static chips (by thresholding out chips of no significant temporal pixel change). We report the averaged MSE and SSIM scores of the demosaicked images in Figures \ref{fig:compare_coded_exposures}(a) and \ref{fig:compare_coded_exposures}(b), respectively.

In the absence of noise, demosaicking performance on NfS is better than on Moving MNIST. Presumably, this is due to the fact that although the images in MNIST are simpler than the natural scenes in NfS, the edges of MNIST are unnaturally sharp and challenging. In the absence of noise, all demosaicking methods were competitive, with FS performing better than CNN in MSE but worse in SSIM.  Unsurprisingly, CNN trained with noise was more robust to increased noise when compared to bilinear interpolation and FS demosaicking (with no explicit noise handling), and noise affected NfS more than Moving MNIST. We conclude that CNN-based demosaicking is best performing overall. However, CNN-based demosaicking was found to be sensitive to calibration errors in practice (which was not modeled in the simulation since it is difficult to obtain ground truth) and thus the real data results were obtained using FS demosaicking.

\subsection{Comparative Study of Coded-Exposure Patterns}

We verify by simulation that the proposed sign-coded FC is robust to noise, compared to the binary-coded one-hot (OH) encoding in \cite{bub2010temporal} as well as the binary-coded pseudo-random (BR) pattern in \cite{reddy2011p2c2}. We made every effort to make this comparison fair and uniform across all coded exposure patterns. The same integer hexagonal lattice TMA sampling was used for spatially multiplexing FC as well as the OH TMA, while BR was implemented by a pseudo-random pattern that is repeated over 16$\times$16 pixel patches as it has foundations in compressive sensing. For reconstruction, all coded-exposure patterns were demosaicked using the same CNN designed with a U-Net architecture~\cite{ronneberger2015u}. It was trained using the exact same training strategies (same architecture, same training patches, noise added in training, the original video sequence as the ground truth data), except for the TMA patterns used to match Hadamard, one-hot, and pseudo-randomize patterns.

The comparative results are summarized in Figures \ref{fig:compare_coded_exposures}(c) and \ref{fig:compare_coded_exposures}(d). With no noise, the three coded-exposure patterns yielded nearly identical performances, with  BR coded exposure slightly outperforming others in NfS. However, even a small amount of noise severely deteriorates the frame reconstruction performance of BR and OH encodings, as evidenced by significantly worse MSE and SSIM scores (despite CNN being trained to handle noise). By contrast, the proposed FC TMA pattern is stable, with graceful performance fall-off over moderate and high levels of noise (especially noticeable in SSIM evaluation). Since the same U-Net demosaicking architecture is used in all coded-exposure patterns, and because nearly identical frame reconstruction quality was achieved in the no-noise scenario, we draw the conclusion that the choice of the coded exposure is the predominant factor in the noise sensitivity of DMD cameras. In fact, the MSE and SSIM scores of the proposed sign-coded FC with FS demosaicking (with \emph{no noise handling}) were in similar performance ranges with the binary-coded pseudo-random exposure with CNN-based demosaicking \emph{trained with noisy data}.

We acknowledge the limitation of this study, in the sense that there are many reconstruction algorithms proposed to date to take advantage of the latest compressive sensing advancements that our study does not necessarily reflect. However, the experiments convincingly support the overall conclusions of this study---that under equal treatment, FC is more noise robust than OH and BR. We expect these conclusions to remain true for further improvements in reconstruction techniques, etc.

\subsection{Real Data Results}
\label{sec:realResults}

\begin{figure*}[htbp]
\centering
\resizebox{\textwidth}{!}{
\begin{tabular}{m{0.01\linewidth}m{0.2\linewidth}m{0.8\linewidth}}
\rotatebox{90}{(a) Chronos} &  &  \includegraphics[width=1\linewidth]{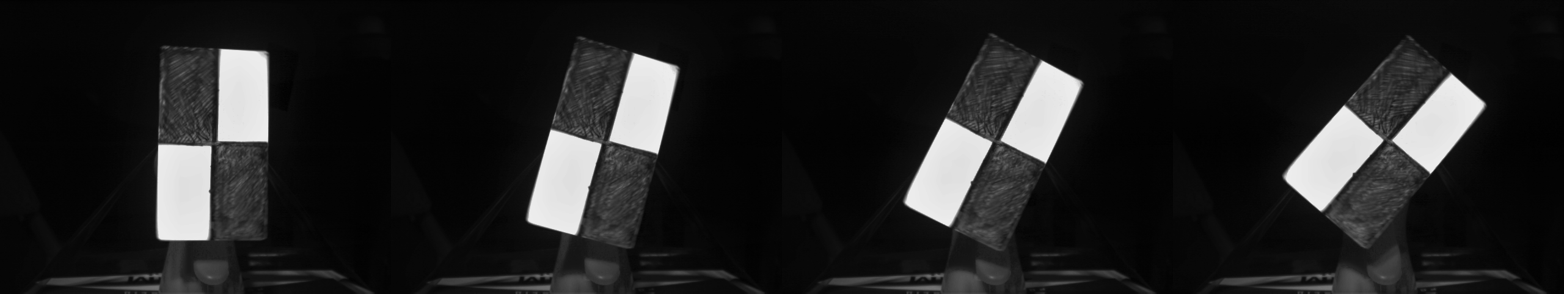}\\
\rotatebox{90}{(b) One Hot} & \includegraphics[width=1\linewidth]{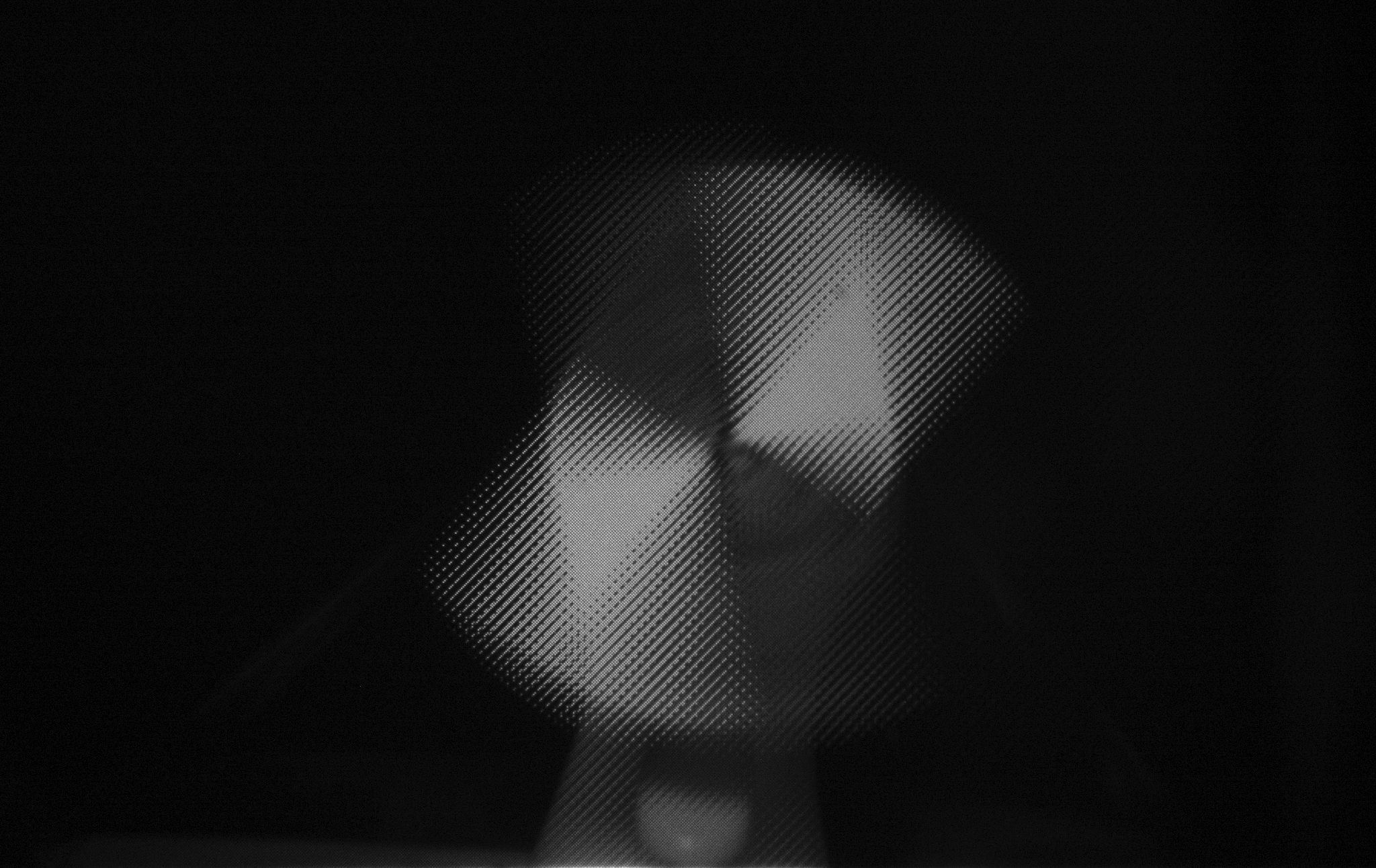} &  \includegraphics[width=1\linewidth]{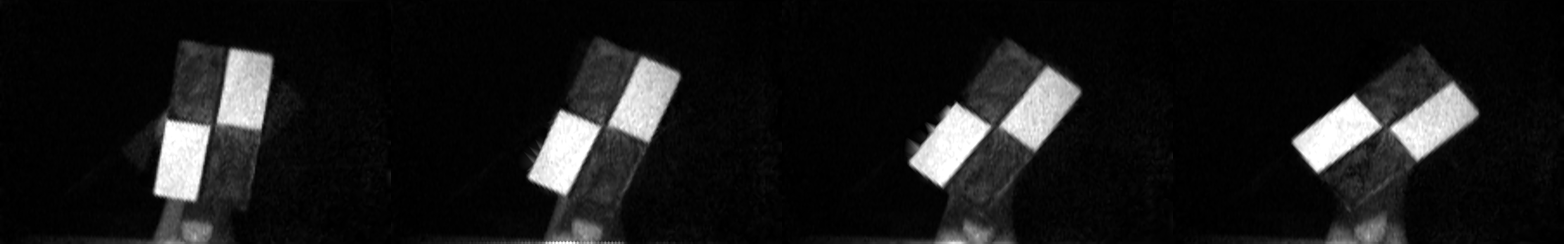}\\
\rotatebox{90}{(c) Sign-Coded FC} & \includegraphics[width=1\linewidth]{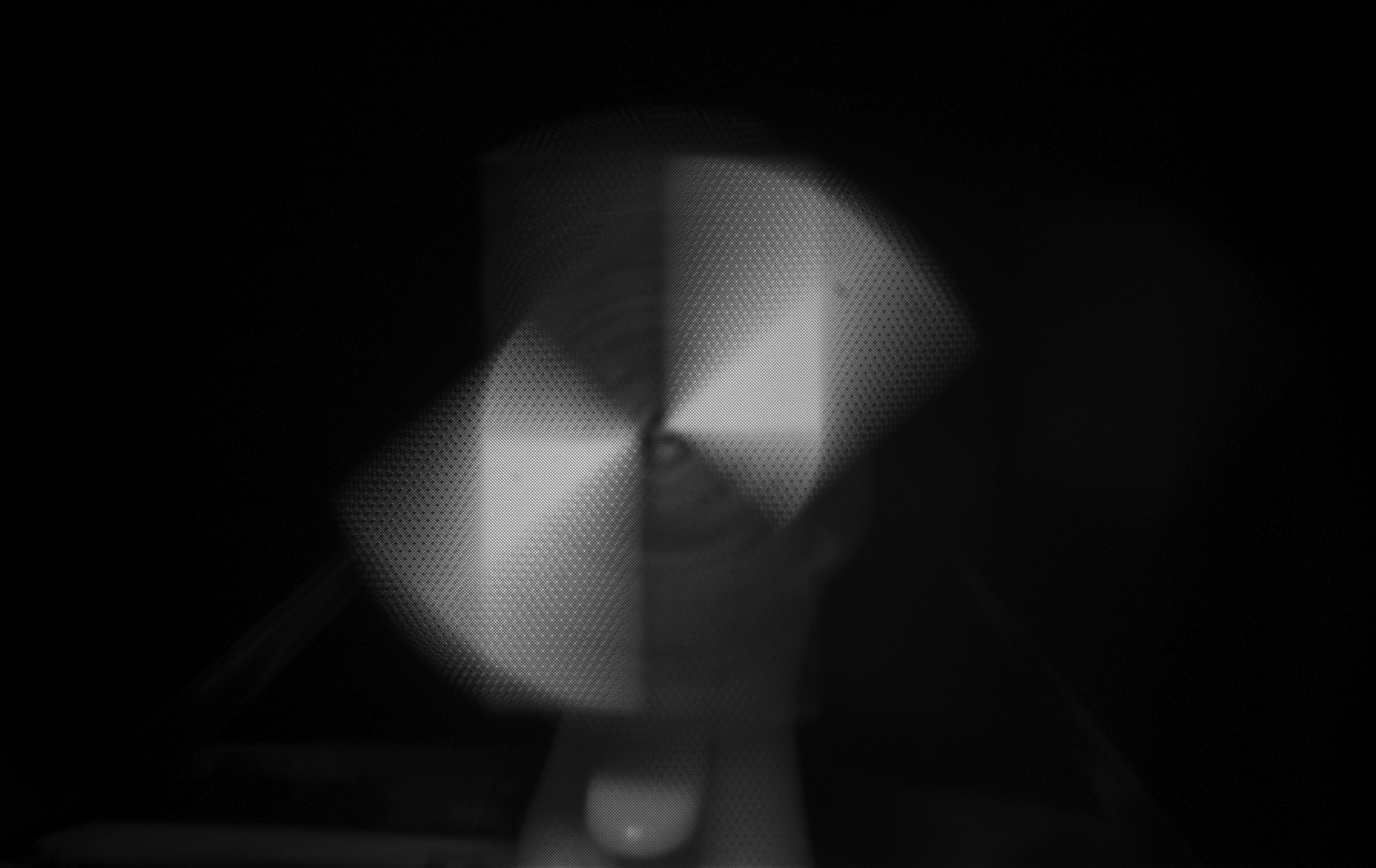} &  \includegraphics[width=1\linewidth]{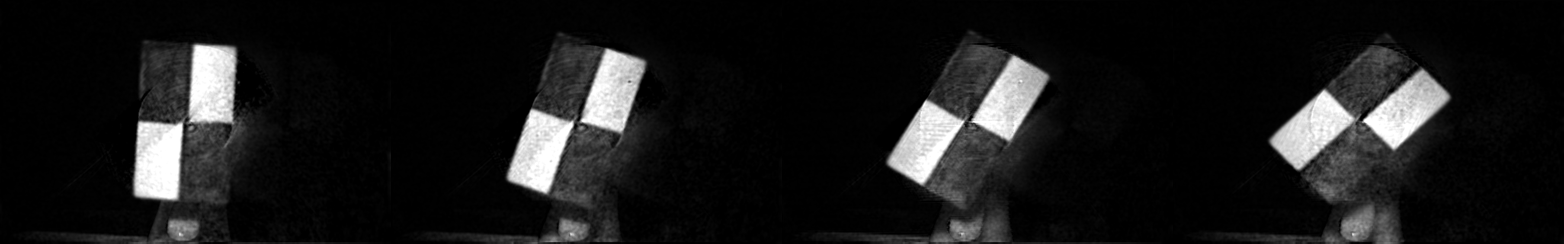}\\
\rotatebox{90}{(d) One Hot} & \includegraphics[width=1\linewidth]{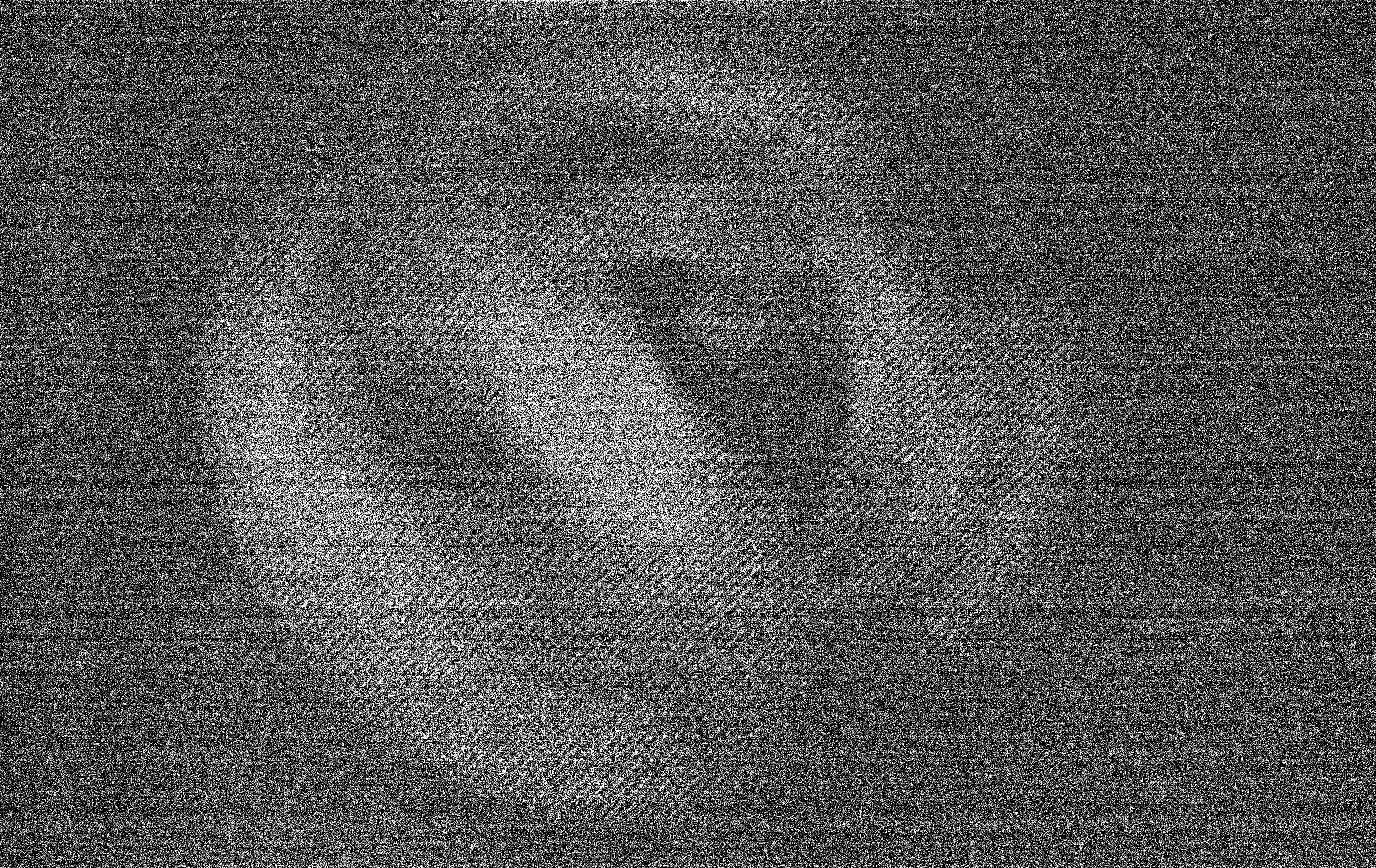} & \includegraphics[width=1\linewidth]{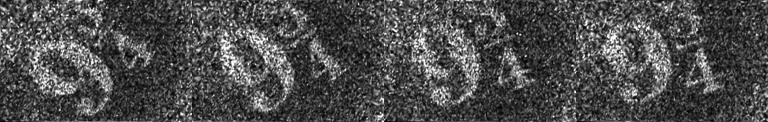}\\
\rotatebox{90}{(e) Sign-Coded FC} & \includegraphics[width=1\linewidth]{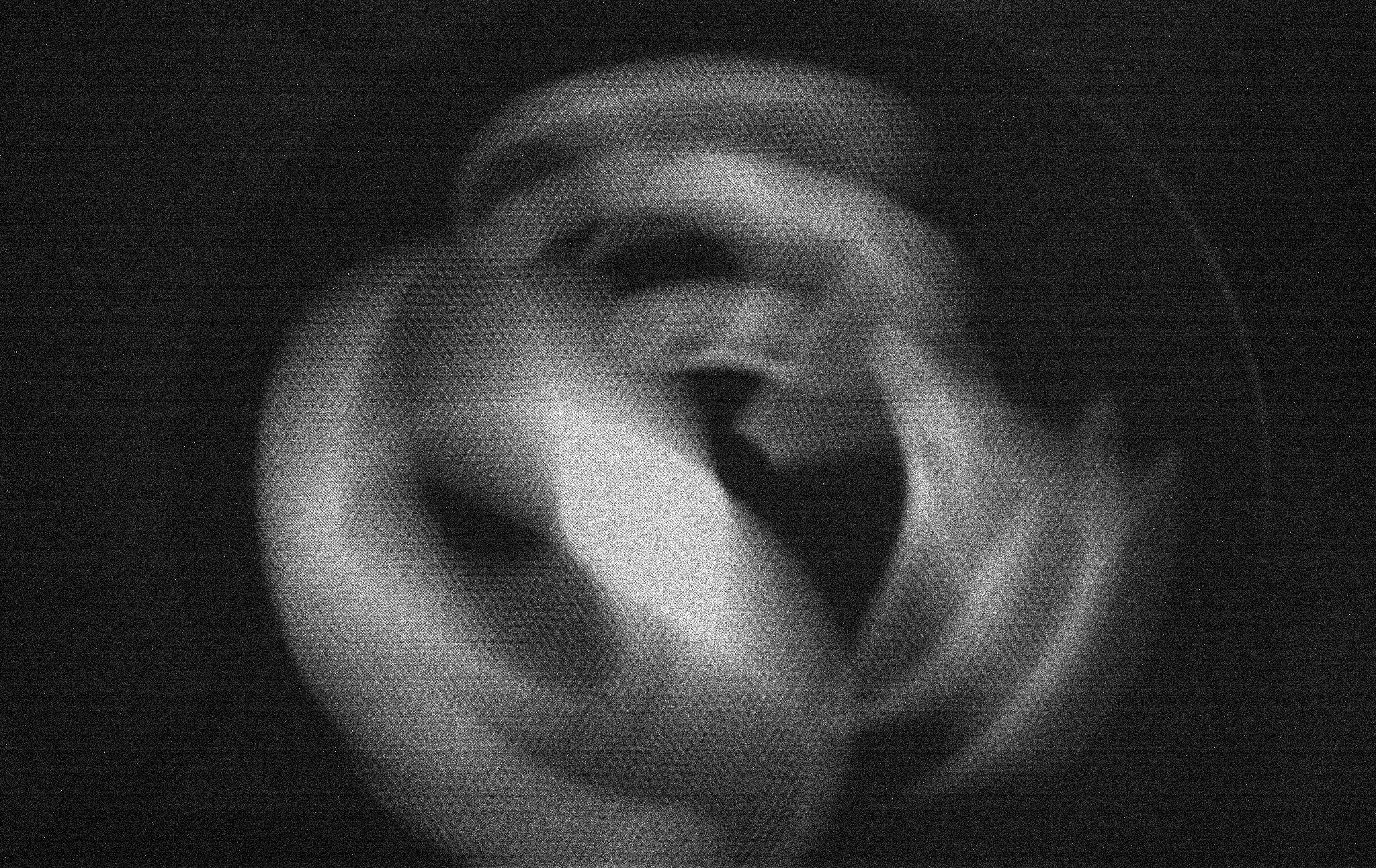} &  \includegraphics[width=1\linewidth]{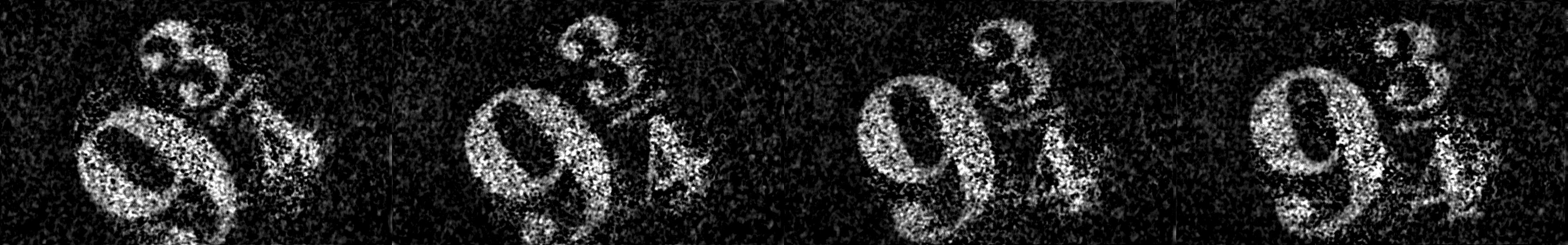}\\
& \hspace{0.8cm}  Raw TMA  Image & \hspace{4.0cm} Four frames (reconstructed from TMA image for (b)-(e))\\
\end{tabular}}
\caption{Hardware prototype evaluation, with DMD operating at \SI{4}{\kilo\hertz} and intensity cameras at \SI{250}{\hertz}. We recorded patterned index cards rotating at \SI{1890}{r\per\minute}---approximately \SI{45}{\degree} travel over the (\SI{4}{\milli\second}) integration. Showing four of the 16 reconstructed \SI{4}{\kilo\hertz} frames. For benchmarking, the same rotating index card was captured by hardware prototype using binary-coded one-hot exposure pattern (same TMA multiplexing), as well as by a conventional high-speed camera (Chronos 1.4). (a-c) When the scene is well-lit (DC studio lights), all reconstruction quality is high. (d-e) In low-light conditions (ambient light), the FC reconstruction is far more robust to noise than one-hot. Note that we adjusted the camera gain so that each of the captured raw TMA images was properly exposed. The one-hot captures needed approximately 8.5 times higher analog gain compared to FC, increasing noise as evidenced by (d).}
\label{fig:fanImages}
\end{figure*}

To test the hardware prototype of the proposed \textbf{Sign-Coded FC Design \#2}, a patterned index card attached to a fan rotating at 1890 revolutions per second was recorded using the FC prototype. With DMD and camera operating at \SI{4}{\kilo\hertz} and \SI{250}{\hertz}, respectively, the card rotates approximately 45${}^o$ over the duration of 16 high-speed frame integration (4ms). For real data experiments, we used the frequency selection demosaicking method described in Section \ref{sec:tfa}. Note that real-time demosaicking at 250Hz is already achievable with today's computational hardware, and is even implemented in resource-limited devices, such as smartphones. 

We also provide two comparisons in hardware. First, we repeat this hardware experiment using one-hot (OH) encoding \cite{bub2010temporal}. We used the same spatial multiplexing pattern (i.e.~integer hexagonal lattice in our TMA) processed by the identical frequency selection demosaicking technique \cite{dubois2005frequency}. Second, we also recorded the same index card with Chronos v1.4, a high-speed camera, set to $640\times480$ resolution at \SI{4}{\kilo\hertz}. The FC and one-hot generated significantly less data (\SI{254}{\kilo\byte}) than the high-speed camera (\SI{4.8}{\mega\byte})---over $18\times$ compression. We adjusted the analog gain of the cameras so that the captured raw TMA images were properly exposed for each capture. Due to the differences in light efficiency, the one-hot captures needed approximately 8.5 times higher analog gain compared to FC.

In our first test, the index card was illuminated using two extremely bright studio lights with DC power supplies (a typical setup for high-speed imaging). The reconstruction from the FC camera shown in Figure~\ref{fig:fanImages}(c) preserves sharp edges with high contrast, and almost matches in quality to the Chronos output in Figure \ref{fig:fanImages}(a) despite the data volume that is orders of magnitude smaller. The OH result in Figure \ref{fig:fanImages}(b) looks as good as our FC reconstruction in Figure \ref{fig:fanImages}(c), matching our modeling under well-lit conditions.

In the second test, the index card was illuminated by natural ambient room light. The analog gain of one-hot encoding in this low-light environment needed to be boosted to the point that the noise in the captured raw TMA image became apparent, due to the fact that the DMD mirrors are active only 1/16 of the time. This significantly degraded the image quality of the one-hot encoding reconstructions in Figure \ref{fig:fanImages}(d). By contrast, the FC camera has significantly better noise performance, thanks to the improved light efficiency, lower analog gain, and the low condition number to avoid noise amplification. In the reconstruction shown in Figure \ref{fig:fanImages}(e), it is possible to distinguish even the smallest edge details.

%% file: 7_conclusion.tex
\section{Conclusion}

We proposed a sign-coded Fourier camera (FC), a novel spatial-temporal light modulator configuration used to carry out coded exposure aimed at in-hardware compression of high-framerate video with minimal noise amplification during frame reconstruction. Specifically, we leverage the orthogonality of the Hadamard transform to encode the temporal evolution of the high-speed image signal, yielding a very low overall condition number compared to the conventional binary randomized encoding pattern. The time modulation array pattern spatially multiplexes Walsh (Hadamard basis) functions, and demosaicking is applied to the captured sensor to yield full-resolution high-speed video frames. We experimentally demonstrated improved robustness to noise over existing binary-coded exposure patterns. The hardware prototype of the sign-coded Fourier camera to operate at 4kHz with only ambient lighting, limited only by the DMD hardware used and not by the proposed sign-coded FC design.

There are several limitations to the prototype design used to implement the FC design. First, the mirror beamsplitter used creates a ghost image (i.e. spatially shifted) due to reflection off the second glass surface. Ghost artifacts could be mitigated by replacing the plate beamsplitter (Figure~\ref{fig:cameraClose}) with a cube beamsplitter (Figure~\ref{fig:camera_design}(c)). Second, light from a single DMD mirror cannot be perfectly focused onto a single AC camera pixel due to scattering. Although the point spread function measured with our optical setup suggests that neighboring coded-exposure patterns suffer from crosstalk, it can be almost completely bypassed by using an on-chip SLM~\cite{luo2019cmos}.

%% file: 0_main.bbl
\begin{thebibliography}{10}
\providecommand{\url}[1]{#1}
\csname url@samestyle\endcsname
\providecommand{\newblock}{\relax}
\providecommand{\bibinfo}[2]{#2}
\providecommand{\BIBentrySTDinterwordspacing}{\spaceskip=0pt\relax}
\providecommand{\BIBentryALTinterwordstretchfactor}{4}
\providecommand{\BIBentryALTinterwordspacing}{\spaceskip=\fontdimen2\font plus
\BIBentryALTinterwordstretchfactor\fontdimen3\font minus
  \fontdimen4\font\relax}
\providecommand{\BIBforeignlanguage}[2]{{%
\expandafter\ifx\csname l@#1\endcsname\relax
\typeout{** WARNING: IEEEtran.bst: No hyphenation pattern has been}%
\typeout{** loaded for the language `#1'. Using the pattern for}%
\typeout{** the default language instead.}%
\else
\language=\csname l@#1\endcsname
\fi
#2}}
\providecommand{\BIBdecl}{\relax}
\BIBdecl

\bibitem{bub2010temporal}
G.~Bub, M.~Tecza, M.~Helmes, P.~Lee, and P.~Kohl, ``Temporal pixel multiplexing
  for simultaneous high-speed, high-resolution imaging,'' \emph{Nature
  methods}, vol.~7, no.~3, pp. 209--211, 2010.

\bibitem{reddy2011p2c2}
D.~Reddy, A.~Veeraraghavan, and R.~Chellappa, ``P2c2: Programmable pixel
  compressive camera for high speed imaging,'' in \emph{CVPR 2011}.\hskip 1em
  plus 0.5em minus 0.4em\relax IEEE, 2011, pp. 329--336.

\bibitem{lincoln2017enhancing}
J.~Lincoln, ``Enhancing physics demos using iphone slow motion,'' \emph{The
  Physics Teacher}, vol.~55, no.~9, pp. 588--589, 2017.

\bibitem{jozwik2016industrial}
J.~J{\'o}zwik, D.~Ostrowski, P.~Jarosz, and D.~Mika, ``Industrial robot
  repeatability testing with high speed camera phantom v2511,'' \emph{Advances
  in Science and Technology Research Journal}, vol.~10, no.~32, 2016.

\bibitem{wang2020single}
P.~Wang, J.~Liang, and L.~V. Wang, ``Single-shot ultrafast imaging attaining 70
  trillion frames per second,'' \emph{Nature Communications}, vol.~11, no.~1,
  pp. 1--9, 2020.

\bibitem{raskar2006coded}
R.~Raskar, A.~Agrawal, and J.~Tumblin, ``Coded exposure photography: motion
  deblurring using fluttered shutter,'' in \emph{ACM SIGGRAPH 2006 Papers},
  2006, pp. 795--804.

\bibitem{holloway2012flutter}
J.~Holloway, A.~C. Sankaranarayanan, A.~Veeraraghavan, and S.~Tambe, ``Flutter
  shutter video camera for compressive sensing of videos,'' in \emph{2012 IEEE
  International Conference on Computational Photography (ICCP)}.\hskip 1em plus
  0.5em minus 0.4em\relax IEEE, 2012, pp. 1--9.

\bibitem{kappal2009illustrating}
S.~J. Kappal and S.~G. Narasimhan, ``Illustrating motion through dlp
  photography,'' in \emph{2009 IEEE Computer Society Conference on Computer
  Vision and Pattern Recognition Workshops}.\hskip 1em plus 0.5em minus
  0.4em\relax IEEE, 2009, pp. 9--16.

\bibitem{nayar2004programmable}
S.~K. Nayar, V.~Branzoi, and T.~E. Boult, ``Programmable imaging using a
  digital micromirror array,'' in \emph{Proceedings of the 2004 IEEE Computer
  Society Conference on Computer Vision and Pattern Recognition, 2004. CVPR
  2004.}, vol.~1.\hskip 1em plus 0.5em minus 0.4em\relax IEEE, 2004, pp. I--I.

\bibitem{luo2019cmos}
Y.~Luo, J.~Jiang, M.~Cai, and S.~Mirabbasi, ``Cmos computational camera with a
  two-tap coded exposure image sensor for single-shot spatial-temporal
  compressive sensing,'' \emph{Optics express}, vol.~27, no.~22, pp.
  31\,475--31\,489, 2019.

\bibitem{deng2019sinusoidal}
C.~Deng, Y.~Zhang, Y.~Mao, J.~Fan, J.~Suo, Z.~Zhang, and Q.~Dai, ``Sinusoidal
  sampling enhanced compressive camera for high speed imaging,'' \emph{IEEE
  transactions on pattern analysis and machine intelligence}, 2019.

\bibitem{liu2013efficient}
D.~Liu, J.~Gu, Y.~Hitomi, M.~Gupta, T.~Mitsunaga, and S.~K. Nayar, ``Efficient
  space-time sampling with pixel-wise coded exposure for high-speed imaging,''
  \emph{IEEE transactions on pattern analysis and machine intelligence},
  vol.~36, no.~2, pp. 248--260, 2013.

\bibitem{mochizuki2016single}
F.~Mochizuki, K.~Kagawa, S.-i. Okihara, M.-W. Seo, B.~Zhang, T.~Takasawa,
  K.~Yasutomi, and S.~Kawahito, ``Single-event transient imaging with an
  ultra-high-speed temporally compressive multi-aperture cmos image sensor,''
  \emph{Optics Express}, vol.~24, no.~4, pp. 4155--4176, 2016.

\bibitem{yuan2021snapshot}
X.~Yuan, D.~J. Brady, and A.~K. Katsaggelos, ``Snapshot compressive imaging:
  Theory, algorithms, and applications,'' \emph{IEEE Signal Processing
  Magazine}, vol.~38, no.~2, pp. 65--88, 2021.

\bibitem{liu2018rank}
Y.~Liu, X.~Yuan, J.~Suo, D.~J. Brady, and Q.~Dai, ``Rank minimization for
  snapshot compressive imaging,'' \emph{IEEE transactions on pattern analysis
  and machine intelligence}, vol.~41, no.~12, pp. 2990--3006, 2018.

\bibitem{zakharov2019few}
E.~Zakharov, A.~Shysheya, E.~Burkov, and V.~Lempitsky, ``Few-shot adversarial
  learning of realistic neural talking head models,'' in \emph{Proceedings of
  the IEEE International Conference on Computer Vision}, 2019, pp. 9459--9468.

\bibitem{jin2018learning}
M.~Jin, G.~Meishvili, and P.~Favaro, ``Learning to extract a video sequence
  from a single motion-blurred image,'' in \emph{Proceedings of the IEEE
  Conference on Computer Vision and Pattern Recognition}, 2018, pp. 6334--6342.

\bibitem{goodfellow2014generative}
I.~Goodfellow, J.~Pouget-Abadie, M.~Mirza, B.~Xu, D.~Warde-Farley, S.~Ozair,
  A.~Courville, and Y.~Bengio, ``Generative adversarial nets,'' in
  \emph{Advances in neural information processing systems}, 2014, pp.
  2672--2680.

\bibitem{jiang2018super}
H.~Jiang, D.~Sun, V.~Jampani, M.-H. Yang, E.~Learned-Miller, and J.~Kautz,
  ``Super slomo: High quality estimation of multiple intermediate frames for
  video interpolation,'' in \emph{Proceedings of the IEEE Conference on
  Computer Vision and Pattern Recognition}, 2018, pp. 9000--9008.

\bibitem{gallego2019event}
G.~Gallego, T.~Delbruck, G.~Orchard, C.~Bartolozzi, B.~Taba, A.~Censi,
  S.~Leutenegger, A.~Davison, J.~Conradt, K.~Daniilidis \emph{et~al.},
  ``Event-based vision: A survey,'' \emph{arXiv preprint arXiv:1904.08405},
  2019.

\bibitem{baldwin2021time}
R.~Baldwin, R.~Liu, M.~Almatrafi, V.~Asari, and K.~Hirakawa, ``Time-ordered
  recent event (tore) volumes for event cameras,'' \emph{arXiv preprint
  arXiv:2103.06108}, 2021.

\bibitem{rebecq2019events}
H.~Rebecq, R.~Ranftl, V.~Koltun, and D.~Scaramuzza, ``Events-to-video: Bringing
  modern computer vision to event cameras,'' in \emph{Proceedings of the
  IEEE/CVF Conference on Computer Vision and Pattern Recognition}, 2019, pp.
  3857--3866.

\bibitem{hirakawa2008spatio}
K.~Hirakawa and P.~J. Wolfe, ``Spatio-spectral color filter array design for
  optimal image recovery,'' \emph{IEEE transactions on image processing},
  vol.~17, no.~10, pp. 1876--1890, 2008.

\bibitem{dubois2005frequency}
E.~Dubois, ``Frequency-domain methods for demosaicking of bayer-sampled color
  images,'' \emph{IEEE Signal Processing Letters}, vol.~12, no.~12, pp.
  847--850, 2005.

\bibitem{lightcrafter20154500}
D.~LightCrafter, ``4500 evaluation module user’s guide,'' \emph{Texas
  Instruments Inc}, pp. 560--590, 2015.

\bibitem{srivastava2015unsupervised}
N.~Srivastava, E.~Mansimov, and R.~Salakhudinov, ``Unsupervised learning of
  video representations using lstms,'' in \emph{International conference on
  machine learning}.\hskip 1em plus 0.5em minus 0.4em\relax PMLR, 2015, pp.
  843--852.

\bibitem{galoogahi2017need}
H.~K. Galoogahi, A.~Fagg, C.~Huang, D.~Ramanan, and S.~Lucey, ``Need for speed:
  A benchmark for higher frame rate object tracking,'' \emph{arXiv preprint
  arXiv:1703.05884}, 2017.

\bibitem{ronneberger2015u}
O.~Ronneberger, P.~Fischer, and T.~Brox, ``U-net: Convolutional networks for
  biomedical image segmentation,'' in \emph{International Conference on Medical
  image computing and computer-assisted intervention}.\hskip 1em plus 0.5em
  minus 0.4em\relax Springer, 2015, pp. 234--241.

\bibitem{he2015delving}
K.~He, X.~Zhang, S.~Ren, and J.~Sun, ``Delving deep into rectifiers: Surpassing
  human-level performance on imagenet classification,'' in \emph{Proceedings of
  the IEEE international conference on computer vision}, 2015, pp. 1026--1034.

\end{thebibliography}
